\documentclass[epj]{svjour}

\usepackage{graphics}
\usepackage{epsfig}

\journalname{The European Physical Journal B}

\newcommand{\be}{\begin{equation}}
\newcommand{\ee}{\end{equation}}
\newcommand{\bea}{\begin{eqnarray}}
\newcommand{\eea}{\end{eqnarray}}
\newcommand{\bfig}{\begin{figure}}
\newcommand{\efig}{\end{figure}}
\newcommand{\bc}{\begin{center}}
\newcommand{\ec}{\end{center}}
\newcommand{\btab}{\begin{tabular}}
\newcommand{\etab}{\end{tabular}}
\newcommand{\dr}{\partial}

\let\oldepsilon\epsilon
\let\epsilon\varepsilon
\let\varepsilon\oldepsilon
\let\oldphi\phi
\let\phi\varphi
\let\varphi\oldphi
\def\EQ{\begin{equation}}
\def\EN{\end{equation}}
\def\EQA{\begin{eqnarray}}
\def\ENA{\end{eqnarray}}
\renewcommand{\deg}{{}^{\circ}}
\newcommand{\ddt}{\partial_{t}}
\newcommand{\ddx}{\partial_{x}}

\newcommand{\nab}{\vec \nabla}

\begin{document}

\title{Selection of dune shapes and velocities.\\
Part 1: Dynamics of sand, wind and barchans.}

\author{Bruno Andreotti\inst{1} \and Philippe Claudin\inst{2} \and
St\'ephane Douady\inst{1}}

\institute{
Laboratoire de Physique Statistique de l'Ecole Normale
Sup\'erieure, 24 rue Lhomond, 75231 Paris Cedex 05, France.
\and
Laboratoire des Milieux D\'esordonn\'es et H\'et\'erog\`enes (UMR 7603),
4 place Jussieu - case 86, 75252 Paris Cedex 05, France.}

\date{\today}

\abstract{Almost fifty years of investigations of barchan dunes
morphology and dynamics is reviewed, with emphasis on the physical
understanding of these objects.  The characteristics measured on the
field (shape, size, velocity) and the physical problems they rise are
presented.  Then, we review the dynamical mechanisms explaining the
formation and the propagation of dunes.  In particular a complete and
original approach of the sand transport over a flat sand bed is
proposed and discussed.  We conclude on open problems by outlining
future research directions.
}

\authorrunning{B. Andreotti, P. Claudin and S. Douady}

\titlerunning{Selection of barchan shapes and velocities. Part 1}

\maketitle

%______________________________________________________________________________
\section{Introduction}
After pioneering works by Bagnolds \cite{B41}, the investigation of
dune morphology and dynamics has been the exclusiveness of geologists
and geographists.  For nearly four decades, they have been reporting
field observations about the condition under which the different kinds
of dunes appear as well as measuring their velocity
\cite{B10,F59,C64,LS64,H67}, their shape
\cite{F59,C64,LS64,H67,N66,HMGP78,H87,S90,HH98,SRPH00}, the patterns
they form, the size distribution of sand grains, etc.  More recently
dunes have attracted the interest of physicists
\cite{SRPH00,WG86,NO93,W95,S97,NY98,DAB99,KSH01} motivated by the
poetry of deserts or by the physics they conceal, by the dryness of
saharan countries or by the observation of dunes on Mars.
\bfig[b!]  \bc \epsfxsize=\linewidth \epsfbox{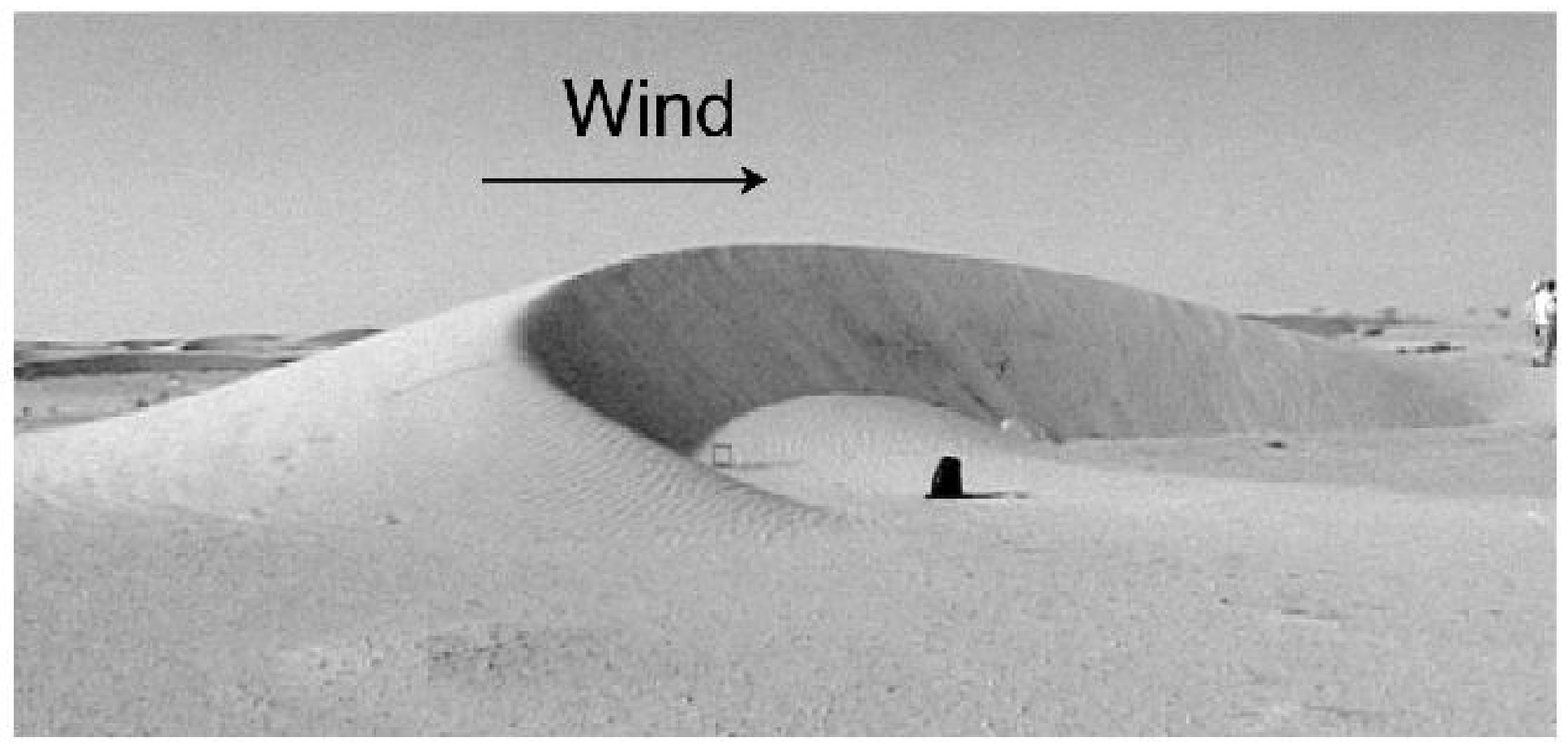} \caption{The
barchan is a dune with a characteristic crescent shape, possibly
isolated.  A small dune ($\simeq 3~m$ high) as that shown
(Mauritania), propagates downwind at one hundred meters per year,
typically.}
\label{Photo}
\ec
\efig

The aim of this paper is to give an overview of this domain.  It is
conceived as a pedagogical review of field observations and of the
dynamical mechanisms important for dune morphogenesis, followed by a
conclusion discussing some of the problems remaining open.  It is
followed by a second part devoted to the derivation of a simple model
predicting the shape and velocity of two-dimensional dunes.
\bfig[b!]  \bc \epsfxsize=\linewidth \epsfbox{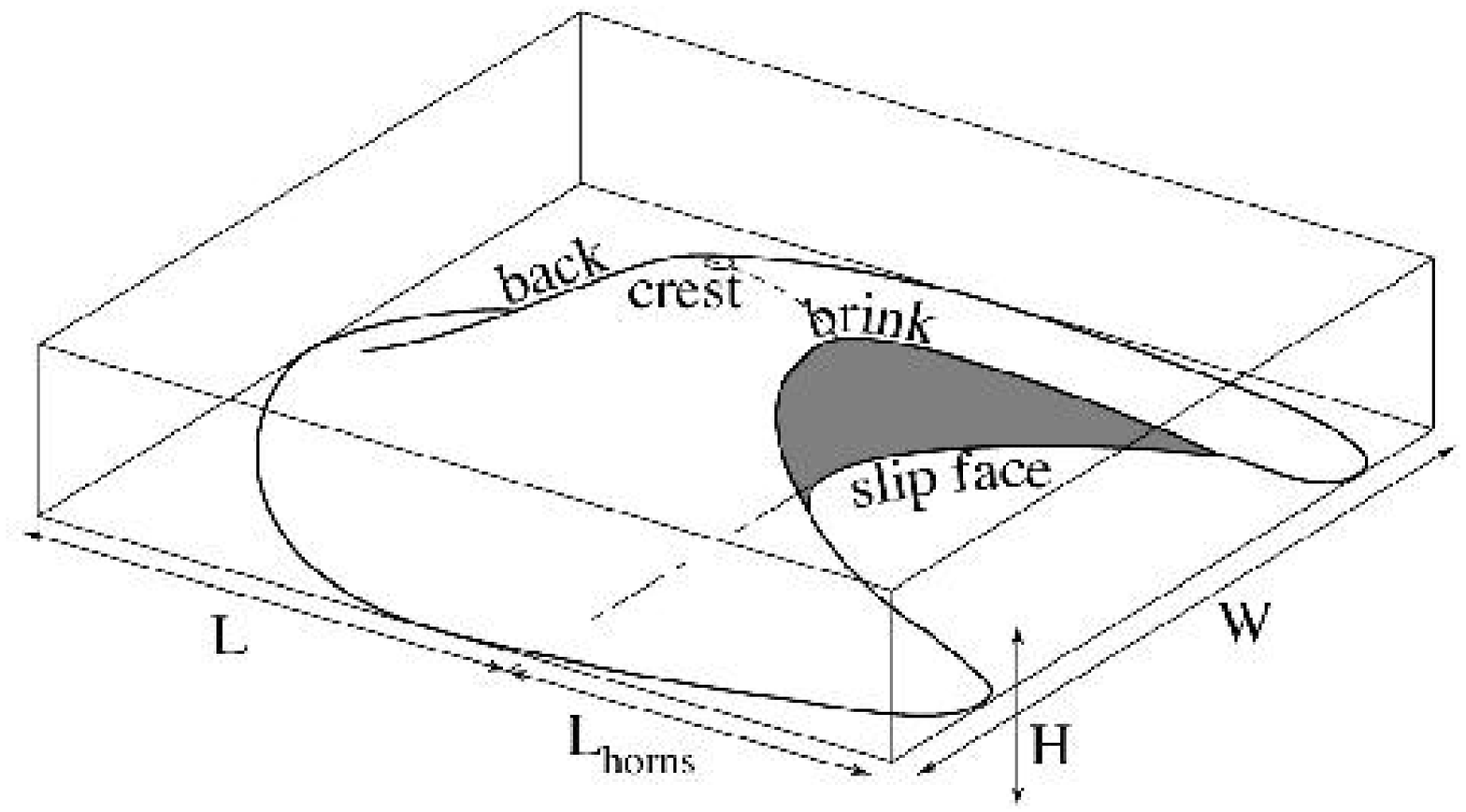}
\caption{Sketch of a barchan dune.  In first approximation, the dune
morphology can be described by four morphologic parameters: the length
$L$, the width $W$, the height $H$ and the horn length $L_{horns}$.}
\label{Schema}
\ec
\efig

In this article, we will mainly focus on the barchan which is the
simplest, and consequently the most studied form of sand dunes.  A
barchan is a crescentic dune as that shown on figure~\ref{Photo},
propagating downwind on a firm soil \cite{B41,CWG93}.  When the
direction of the wind is almost constant, these dunes can maintain a
nearly invariant shape and size ($3-10~m$) for very long times
($1-30~years$) \cite{B41,F59}.  The basic dynamical mechanism
explaining the dune propagation is simple (see figure~\ref{Schema}):
the back of the dune is eroded by the wind; the sand transported in
saltation is deposed at the brink, and is redistributed on the
slip-face by avalanches.  In this picture, a dune keeps the same
amount of sand.  In reality, the global mass evolution results from
the balance between the sand supply at the back and the leak at the
horns.
\bfig[b!]  \bc \epsfxsize=\linewidth \epsfbox{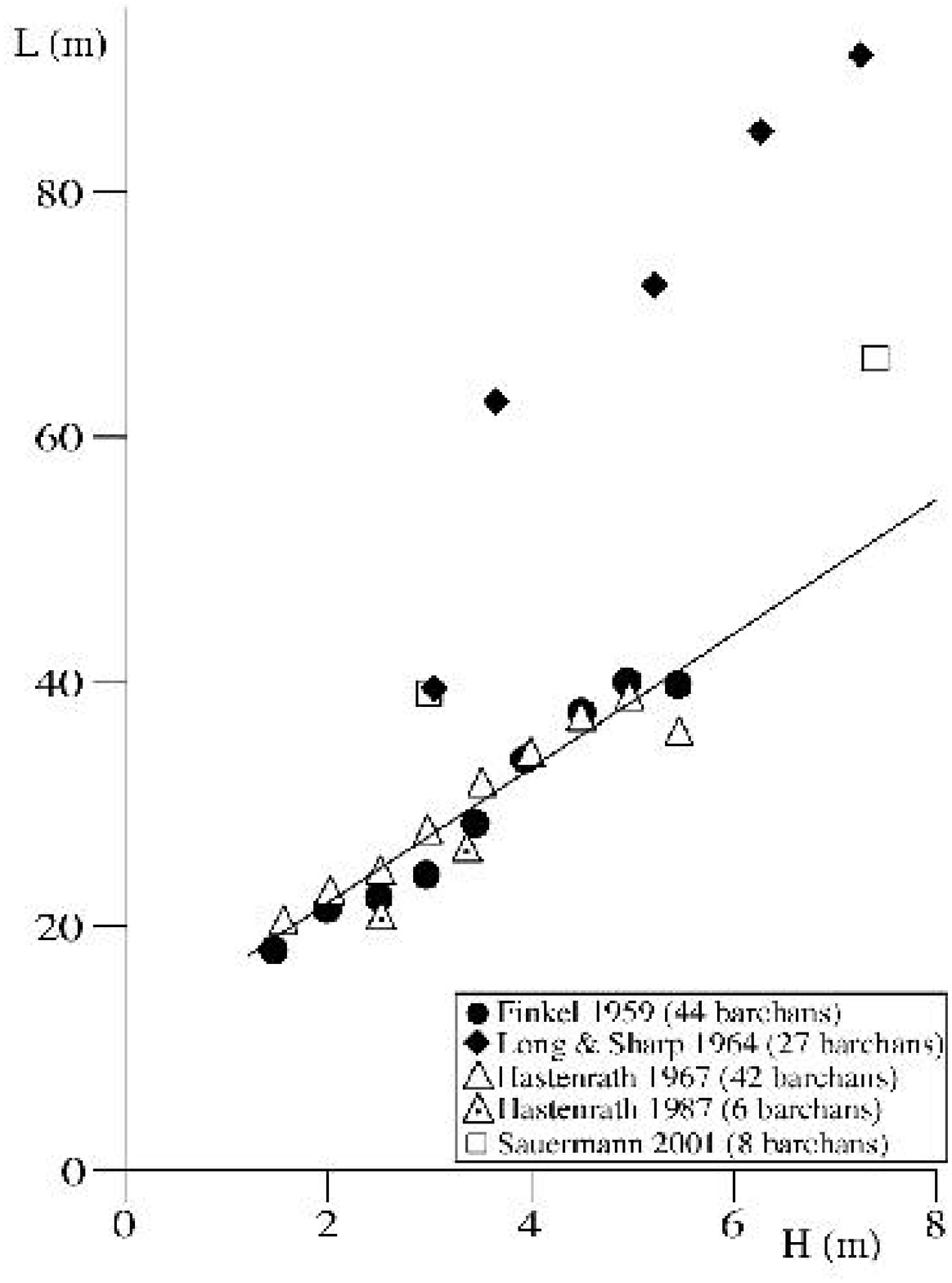}
\caption{Relationship between the dune length $L$ and height $H$
determined from field measurements averaged by ranges of heights.  The
solid line is the best linear fit to the points corresponding to
barchans from the Arequipa region in Southern Peru (Finkel \cite{F59}
and Hastenrath \cite{H67,H87}).}
\label{LofH}
\ec
\efig

Starting from this overview of barchan properties, we give in section
\ref{fieldobs} a short review of the field observations, concerning in
particular the morphology and velocity scaling laws.  The dynamical
mechanisms involved in erosion and sand transport are presented in
section \ref{mechanisms} and the corresponding dimensionless
parameters as well as the scaling laws are discussed.  Finally, in
section \ref{problems}, we sort the facts both validated by
experimental measurements and explained by consistent theories from
the problems remaining partly or totally open up to now.

%______________________________________________________________________________
\section{Field observations}
\label{fieldobs}
\subsection{Morphologic relationships and scale invariance}
The crescent-like shape of the barchan is well known (see
figures~\ref{Photo}~and~\ref{Schema}) and can be characterized by a
few parameters, the length $L$ along the central axis (which is also
the wind direction), the height $H$, the width $W$ and the horn length
$L_{horns}$ (see figure~\ref{Schema}).  Note that in practice, the two
horns have always different lengths due to the fluctuations of wind
direction.  They are thus measured separately and averaged to give
$L_{horns}$.

The region around La Joya in southern Peru is the most documented
barchan field \cite{F59,H67,H87,G99} and that for which measurements are
the most coherent.  Finkel \cite{F59} and later Hastenrath
\cite{H67,H87} have chosen a squared region, a few kilometres width,
and have investigated systematically the morphologic parameters of the
barchans in this perimeter.  Their data exhibit a large statistical
dispersion due to the variations of the control parameters on the
field.  Then, it is not surprising to find a dependence of the dune
shape on the local conditions like the wind regime, the sand supply,
the presence or not of other dunes in the vicinity, the nature of the
soil, or wether the studied dunes have achieved a permanent state or
are in a transient state\ldots
\bfig[t!]  \bc \epsfxsize=\linewidth \epsfbox{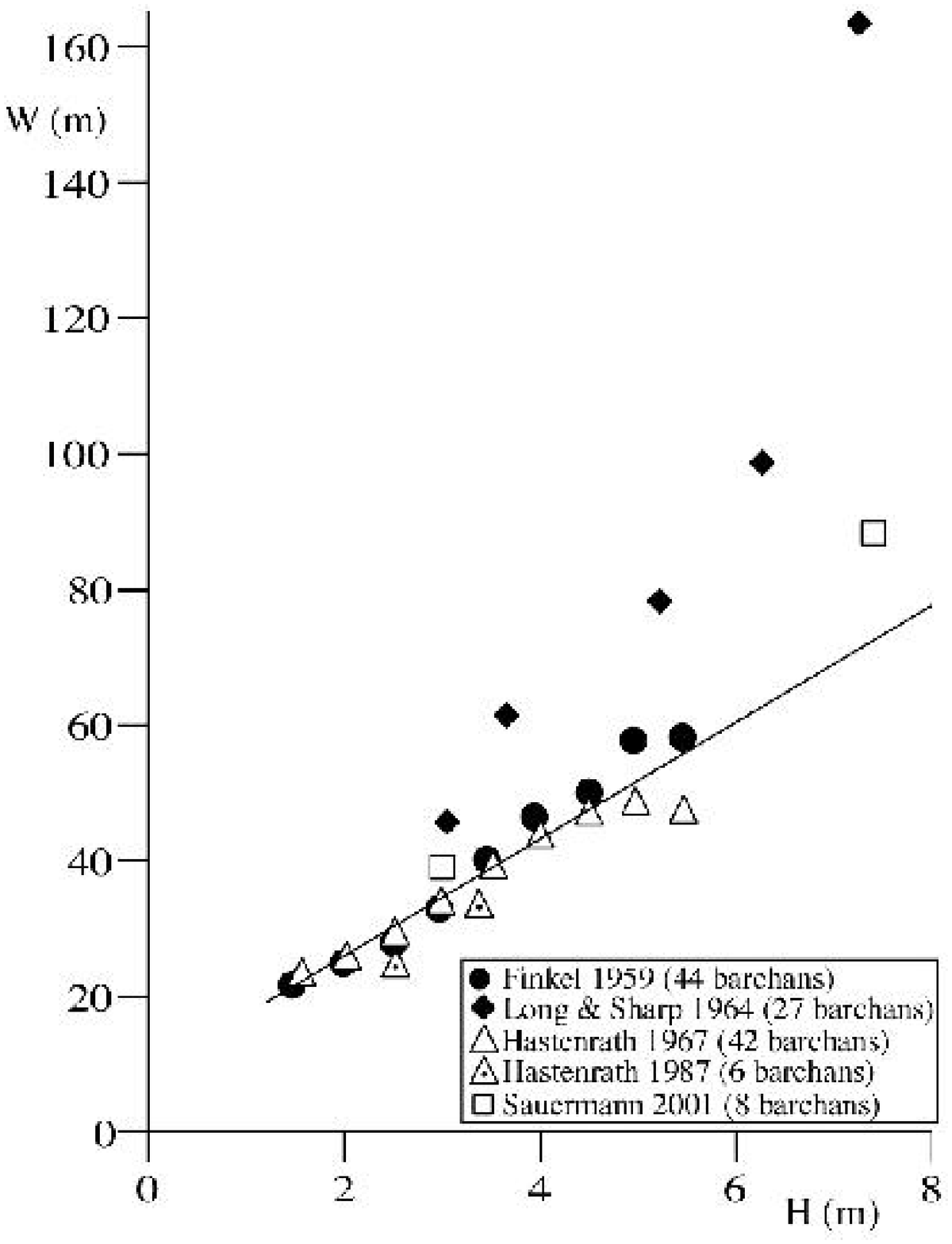}
\caption{Relationship between barchan width $W$ and height $H$
determined from field measurements averaged by ranges of heights.  The
solid line is the best linear fit to the points corresponding to
barchans from the Arequipa region in Southern Peru (Finkel \cite{F59}
and Hastenrath \cite{H67,H87}).}
\label{WofH}
\ec
\efig

Despite this strong dispersion of data, clear linear relationships
between the height $H$, the length $L$ and the width $W$ (see
figure~\ref{Schema}) were found by Finkel \cite{F59} and Hastenrath
\cite{H67}, but with different coefficients ($W=W_0+\rho_W H$ and
$L=L_0+\rho_L H$).  To seek for relationships valid statistically, we
follow the procedure of Finkel \cite{F59} and average all the
measurements by ranges of heights (see
figures~\ref{LofH}~and~\ref{WofH}).  More precisely, we have
considered the barchans between $1~m$ and $2~m$ high and averaged
their length, width and height; this gives one point on
figures~\ref{LofH} and~\ref{WofH}.  The same computation was done for
the barchans between $1.5~m$ and $2.5~m$, between $2~m$ and $3~m$,
etc.  We can reasonably expect the statistical relations to be that
which would have been obtained in controlled, reproducible conditions.

It can be observed that linear relationships between the length $L$,
the width $W$ and the height $H$ are nonetheless recovered but are
actually the same for the three sets of measurements (Finkel
\cite{F59} and Hastenrath \cite{H67,H87}) made in Southern Peru.  The
best linear fits give $\rho_W=8.6$ and $\rho_L=5.5$ corresponding to the
mean proportions of barchans near La Joya.  The most important point is
that $L$, $H$ and $W$ are not proportional.  On the contrary, the
coefficients $L_0=10.8~m$ and $W_0=8.8~m$ {\it are comparable to the
size of the smallest dunes}.  This means that small and large dunes are not
homothetic: dunes of different heights have different shapes.  In
other words, {\it barchans are not scale invariant objects}.  As a
consequence, there should exist (at least) a typical lengthscale
(related to $L_0$ and $W_0$) in the mechanisms leading to the dune
propagation.

There are very few places in the world similar to the Arequipa region
(southern Peru).  They are usually touched by trade winds driven by
oceanic anticyclones.  Because these anticyclones are very stable, the
winds are constant and often very strong.  It is the case with the
Peruvian and Chilean coasts, with coastal Namibia, with the Atlantic
coast of the Sahara from Morocco to Senegal and with the northern
shores of Western Australia.  Morphologic measurements were conducted
in some of the places where there is a constantly prevailing wind
direction and a smooth ground surface: in Mauritania by Coursin
\cite{C64}, in California by Long and Sharp \cite{LS64}, and more
recently in Namibia by Hesp and Hastings \cite{HH98} and in southern
Morocco (former Spanish Sahara) by Sauermann {\it et al.}
\cite{SRPH00}.  In the later case, the authors have not measured the
whole barchans in a given area but selected some which where isolated
and symmetric.  Unfortunately, in these works, the number of barchans
is too small to determine precise morphologic relationships.  But they
are sufficient to indicate that the morphology depends on the dune
field (see figures~\ref{LofH}~and~\ref{WofH}).  Briefly, we know
almost nothing on the parameters important for dunes morphology.  It
can depend on the grain size, the density of dunes, the sand supply,
the wind strength, its changes of direction, etc.

For instance, it has been observed by Allen \cite{A68} that barchans
have a smaller width $W$ and more developed horns $L_{horns}$ under
strong winds than light ones.  But Hastenrath have reported
measurements which indicate a weak dependence of the shape on the wind
strength.  He returned to southern Peru \cite{H87} after a light wind
decade and observed that all the dunes had strongly decreased in size.
He measured a second time the morphologic parameters of $6$ barchans
(dotted triangle on figures~\ref{LofH}~and~\ref{WofH}): the points
are almost on the same line than previously.

Another example is the coincidence (or not) of the brink and the crest
(see figure~\ref{Schema} for a definition of these words).  Hastenrath
\cite{H67} and Sauermann {\it et al.} \cite{SRPH00} have observed that
small dunes present a broad domed convexity around the crest, clearly
separated from the brink as in figure~\ref{Schema}, while large dunes
have the crest straight to the brink as in figure~\ref{Photo}.  This
is represented schematically on figure~\ref{Q=vL}.  However, it had
been reported before \cite{CWG93} that there exist barchans of the
same height in the same dune field presenting alternatively the
separation or the coincidence of the brink and the crest.  The
difference lies perhaps in the selection of barchans to be studied
(isolated or not, symmetrical or not, etc).
\bfig[b!]  \bc \epsfxsize=\linewidth \epsfbox{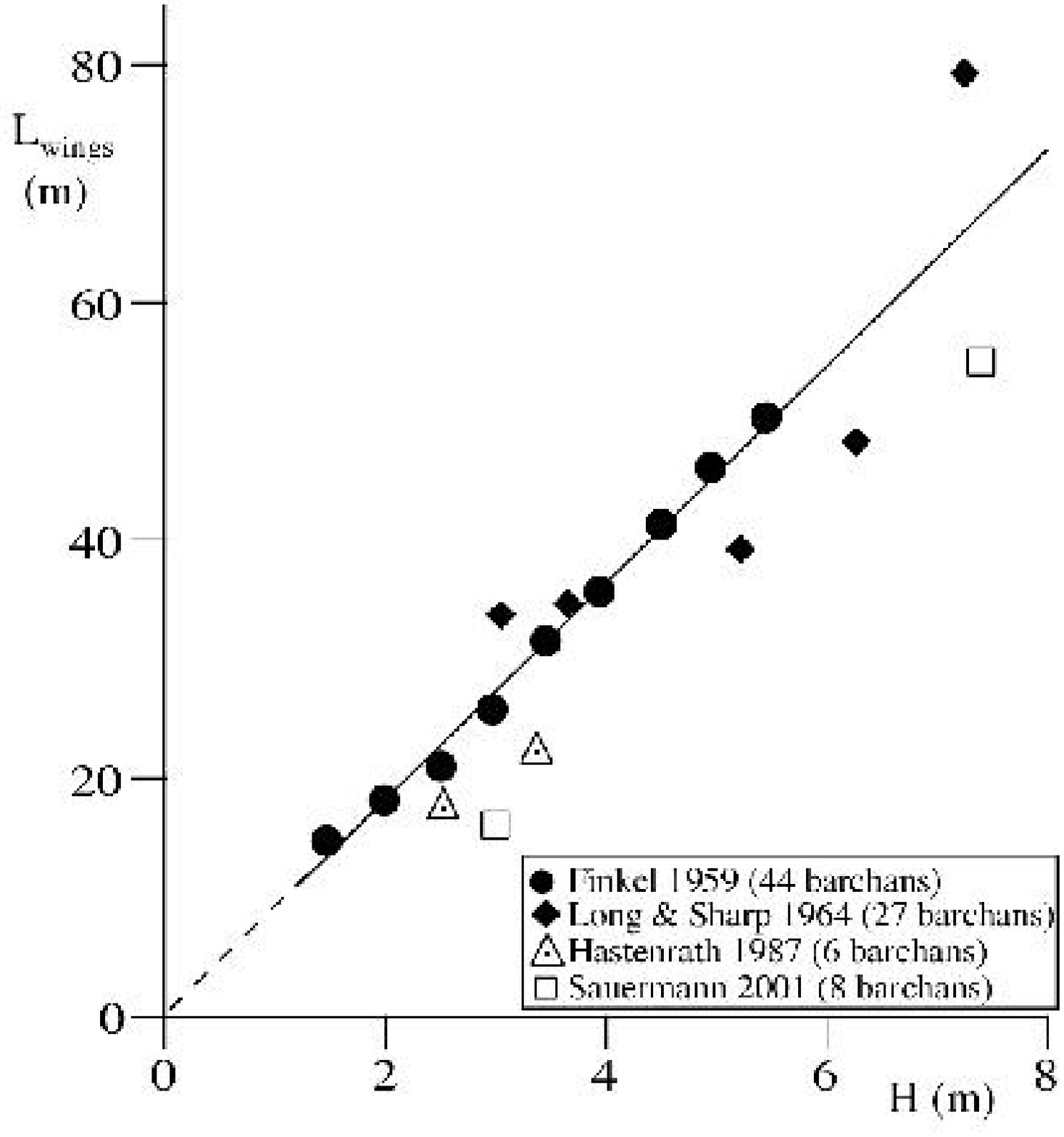}
\caption{Relationships between the horn length $L_{horns}$ and the
width $W$ determined from field measurements averaged by ranges of
heights.  The solid line is the best linear fit to the points
corresponding to barchans from the Arequipa region in Southern Peru
(Finkel \cite{F59} and Hastenrath \cite{H67,H87}).}
\label{LhornsofH}
\ec
\efig

The last morphologic relation between the mean length of the horns
$L_{horns}$ and the barchan height $H$ (figure~\ref{LhornsofH}) is a
nearly perfect relation of proportionality: $L_{horns}\simeq 9.1~H$.
Even more striking, the relation is approximately the same for all the
dune fields measured (in particular La Joya, Peru \cite{F59,H67,H87}
and Imperial Valley, California \cite{LS64}, see
figure~\ref{LhornsofH}).  This suggests that the scaling law of the
horns length could be simpler and more robust than that of the back
dimensions.
\bfig[t!]  \bc \epsfxsize=\linewidth \epsfbox{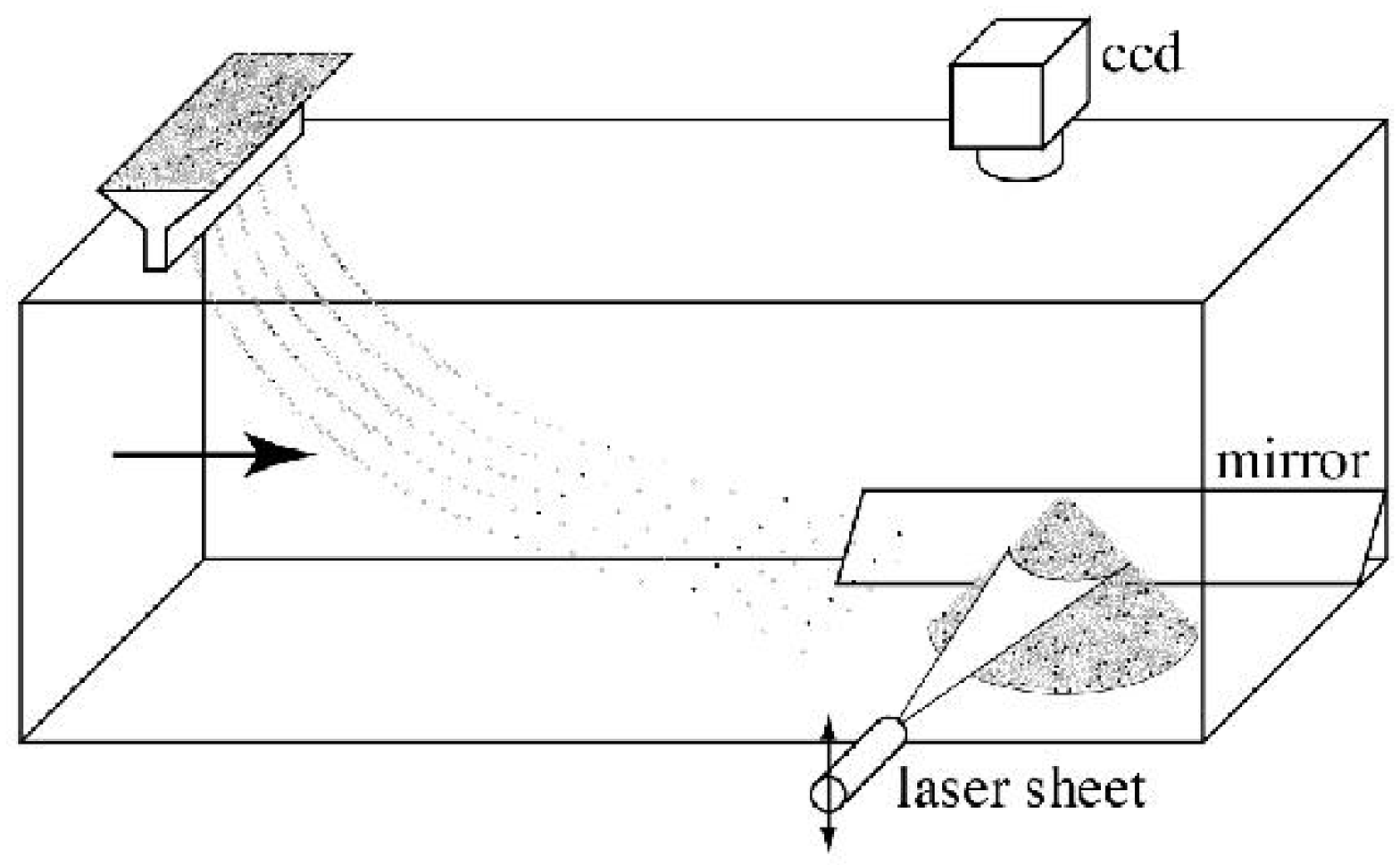}
\caption{To study the time evolution of a sandpile blown by the wind,
the Cemagref wind tunnel ($6~m$ long, $1~m$ large and $1~m$ high) was
used.  Pictures were taken from above using a video camera.  A mirror
was placed at $45 \deg$ to get on the same picture a side view and a
top view of the pile.  The later was enlightened by a lamp and by an
horizontal laser sheet adjustable in height which reveals the
topography.  A tunable sand supply has been add at the top of the
tunnel.  The grains are PVC beads of size $100~\mu m$.  The velocity
(around $6~m/s$ at $2~cm$ above the soil covered by velvet) is chosen
slightly above the threshold of motion of the grains.}
\label{Schema_Soufflerie}
\ec \efig
\begin{figure*}
\epsfxsize=\linewidth
\epsfbox{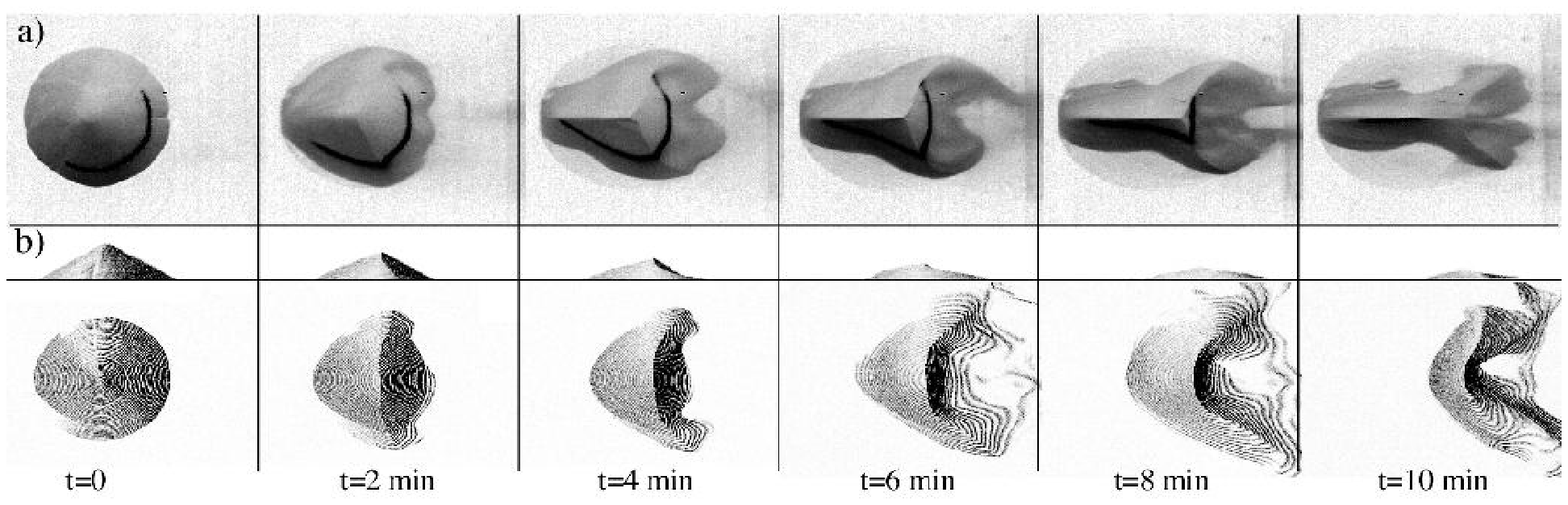} \caption{Time evolution of a sandpile
blown out by a controlled wind without (a) and with (b) a sand supply.
In both cases the sandpile is eroded and disappears after a few
minutes.  Without sand supply, the erosion is localised on the sides
so that a longitudinal brink is created.  With a sand supply, the pile
takes the form of a crescent.  In particular, the back remains smooth
and two horns grow by deposition of grains in reptation.  In case b),
both top an side views are shown.  In particular, the initial height
of the pile ($\simeq 20~cm$) can be seen.}
\label{Film_Soufflerie}
\end{figure*}

\subsection{Minimum size}
\label{MinimumSize}
It is striking to note the absence, on
figures~\ref{LofH},~\ref{WofH}~and~\ref{LhornsofH}, of measured dunes
smaller than $H=1~m$, $W=19~m$ and $L=17.5~m$.  This cut-off is
clearly visible on the dune size histograms measured by Hastenrath
\cite{H67}.  Moreover, if barchans can be very thin sand patches in
the region where they form, they are never lower than, say, $1~m$ in
mature dunes regions.  What appends to a small barchan~?  Related to
this question, there have been several attempts \cite{B41,CWG93,DLG01} to
generate an artificial dune from a small conical sandpile (typically
$10~cm$ to $1~m$ high).

Here we report an experiment made in the Cemagref wind tunnel,
described on figure~\ref{Schema_Soufflerie}.  A conical sandpile
$20~cm$ high is built from a funnel.  It is then eroded by the air
flow with (case b) or without (case a) a sand supply at the beginning
of the wind tunnel (figure~\ref{Film_Soufflerie}).  Whatever the
conditions (even with a sand supply), the pile looses mass inexorably,
and disappears in a few minutes.  This is also what was found in field
experiments \cite{B41,CWG93,DLG01}.

Figure~\ref{Film_Soufflerie} shows the evolution of the sandpile in
the two cases, for a wind velocity chosen slightly above the threshold
of motion of the grains.  In case a), without a sand supply, the
erosion is localised on the sides and the formation of a longitudinal
brink is observed.  In the last steps the side faces become so steep
that avalanches occur.  In case b), the input sand flux is tuned to be
slightly below the saturated flux.  Both the back and the sides are
eroded so that the pile takes the form of a barchan.  In particular,
we observe the progressive formation of two horns due to the lateral
deposition of grains in reptation.

The observation that a sandpile $20~cm$ high disappears whatever the
conditions suggests that barchan dunes have a minimal size.  This was
first noticed by Bagnolds who interpreted the cut-off scale (the
minimum dune size) as the saturation length which will be defined and
discussed in details in section \ref{saturationpara}.  Basically, it
is the length over which the sand carrying increases when the wind
passes from a firm soil to a sand patch.  This interpretation has not
been confirmed so far and deserves further investigations.  Still the
existence of a minimal size rises crucial problems.  First, it means
that no small size dune can be obtained in the air.  Second, it asks a
fundamental question : if a small barchan (or a conical small
sandpile) is unstable and disappears after a short time, how then can
barchans form?

\subsection{Barchans velocity}
\label{velocc}
In many places, barchans velocity has been measured
\cite{F59,B10,C64,LS64,H67,H87,S90}.  Typically, a small dune, $3~m$
high, propagates at a speed ranging from $15~m/year$ to $60~m/year$
while this velocity is between $4~m/year$ and $15~m/year$ for a large
dune, say $15~m$ high.  Even though measurement points are very dispersed
\cite{CWG93}, there is no doubt that small dunes move faster than
large ones.
\bfig[h!]  \bc \epsfxsize=\linewidth \epsfbox{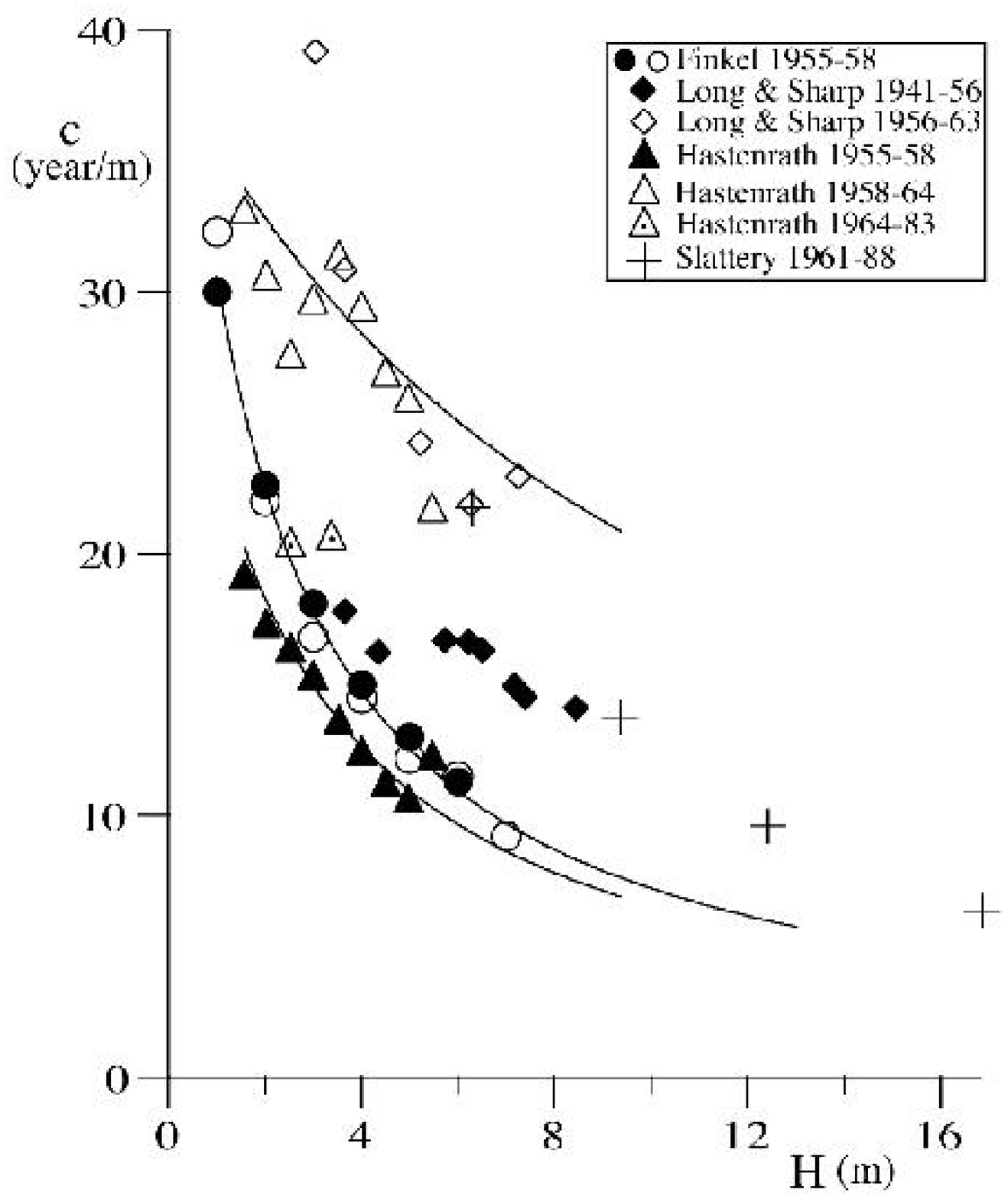}
\caption{Dune velocity, averaged by ranges of height, as a function of
height.  Solid lines correspond to the best fit of the data by a
velocity of the form $c = Q/(H_0+H)$.  The fit is shown for the
measurements in the Arequipa region (Finkel \cite{F59} and Hastenrath
\cite{H67,H87})}
\label{VitBarchan}
\ec
\efig

As for morphologic relationships, we have grouped the measurement
points and averaged the velocity over several barchans.  The resulting
curves are shown on figure~\ref{VitBarchan}.  It can be seen that the
dune velocity is very different from one place to another.  In the
imperial valley, California, dune displacements were measured by Long
\& Sharp \cite{LS64} from 1941 to 1956 (black diamonds), and from 1956
to 1963 (white diamonds).  In Peru, they were measured by Hastenrath
\cite{H67,H87} from 1955 to 1958 (black triangle), from 1958 to 1964
(white triangle), and from 1964 to 1983 (dotted triangle).  From these
two data sets, it can be seen that barchan velocity $c$ strongly
depends on time.  This is probably related to the fact that it
obviously depends on fluctuating parameters like the wind speed or the
sand supply.
\bfig[t!]  \bc \epsfxsize=\linewidth \epsfbox{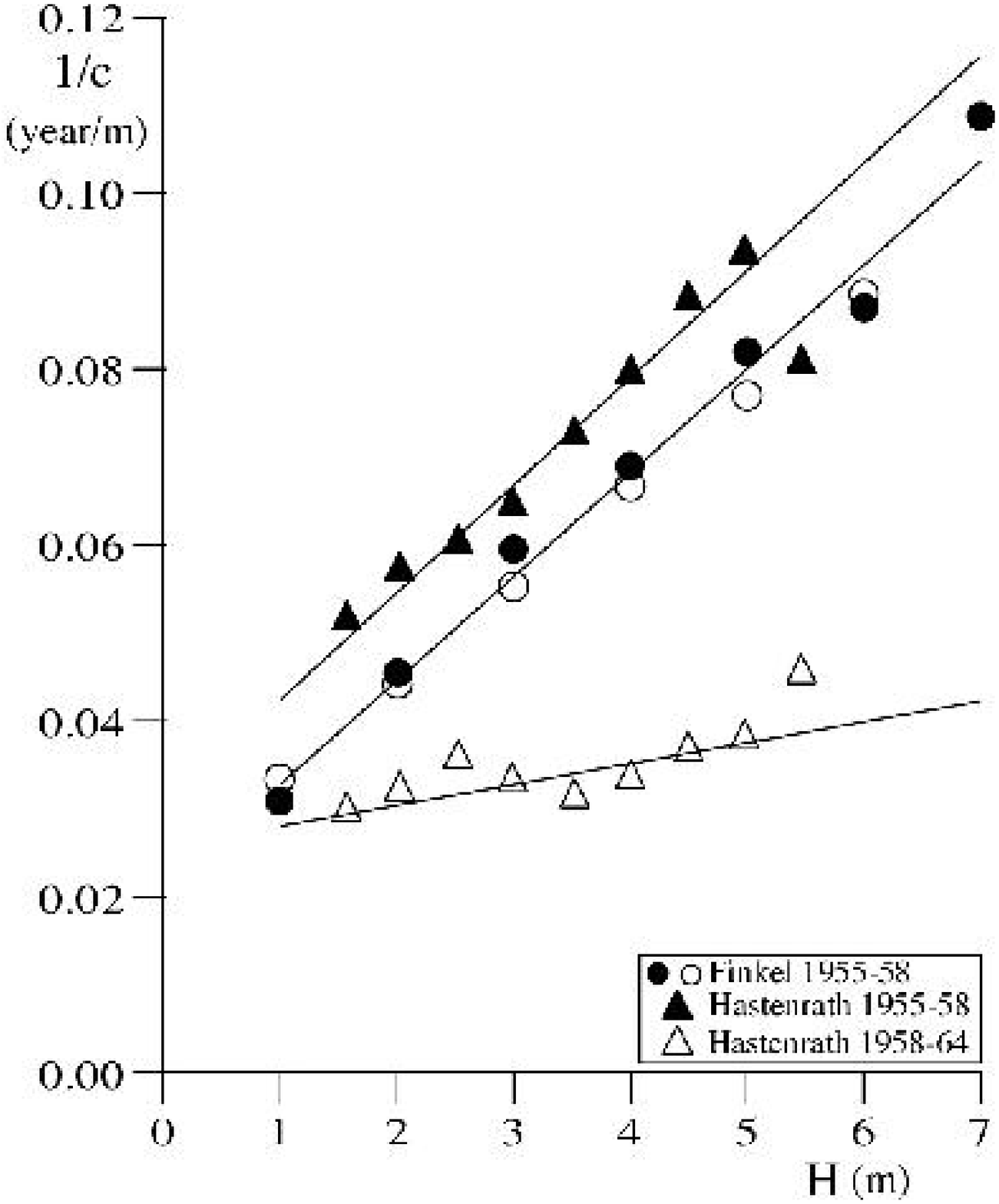}
\caption{Inverse dune velocity $1/c$, averaged by ranges of height, as
a function of height $H$.  Only the measurements made in the region of
La Joya (Southern Peru) are shown.}
\label{UnVitBarchan}
\ec
\efig

\begin{figure*}
\epsfbox{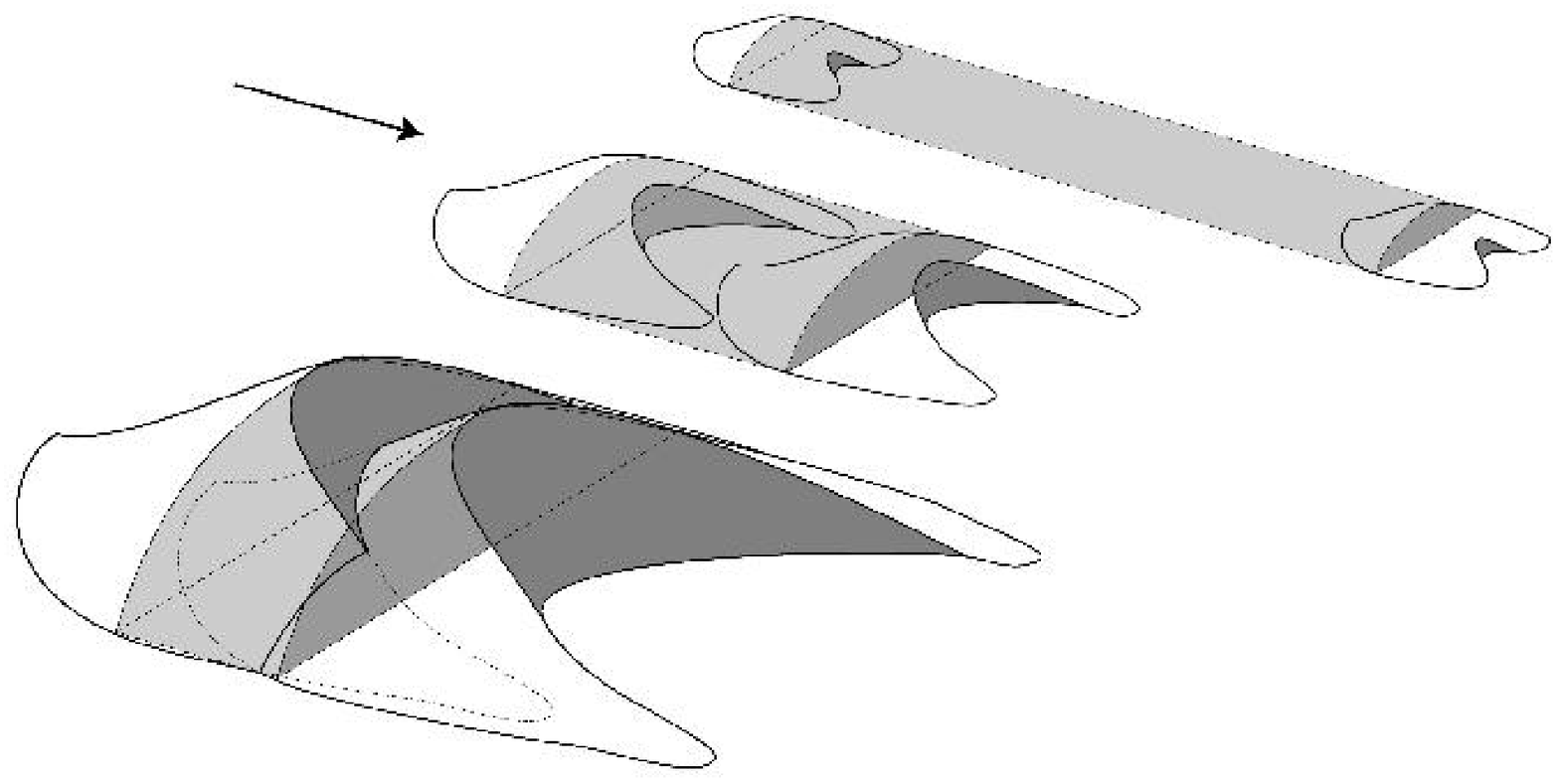} \caption{Visual representation of the relation
between the dune velocity $c$ and its height $H$: three barchans are
represented at initial time and after some time $t$.  It can be seen
that small dunes go faster than large ones.  From the scalings of the
speed $c$ and the width $W$ (see text), we can infer that the area
$W~ct$ swept by the barchan (in light grey) is almost independent of
the dune size.  It can be seen on the large barchan that the back of
the dune is eroded while the slip face and the horns are regions of
sand deposition.
\label{Q=vL}}
\end{figure*}

However, in each dune field and especially in Southern Peru
(figure~\ref{UnVitBarchan}), the relationship between velocity and
size can be reasonably described by:
\be c \simeq \frac{Q}{H_0+H}.
\label{c=Q/H}
\ee
This relation is represented schematically on figure~\ref{Q=vL}.  $Q$,
which is homogeneous to a volumic sand flux, is found to be
approximately equal to $85~m^2/year$ between 1955 and 1958 and to
$425~m^2/year$ between 1958 and 1964, in southern Peru
(figure~\ref{UnVitBarchan}).  $H_0$ is a cut-off height ranging from
$1.8~m$ (Finkel 1955-58) to $10.9~m$ (Hastenrath 1958-64).

>From the propagation speed, we can construct the dune turnover time
$T_{turnover}$ as the time taken by the dune to travel over its own
length:
\be T_{turnover}=\frac{L}{c}\simeq \frac{L (H_0+H)}{Q}. \ee
It is also the typical period of the cycle of motion of a grain of the
dune: erosion from the back, deposition at the top of the slip face,
flowing in an avalanche, at rest below the dune, reappearance at the
back of the dune, etc.  It gives also the typical time needed by the
dune to readjust its whole shape to changes of external conditions
(wind, sand supply\ldots).  Typically, a small dune, $3~m$ high, has a
turnover time between $5~months$ and $2~years$ while it ranges from
$6~years$ to $25~years$ for a large dune, $15~m$ high.  According to
Oulehri \cite{O92}, this large difference of turnover times between
small and large dunes was already known in ancient times by saharan
people.  Barchans were used as cereal lofts or to protect goods from
pillaging.  A small or a large dune was chosen, function of the time
after which they wanted to recover the bundle.

%______________________________________________________________________________
\section{Mechanisms, dimensionless parameters and scaling laws}
\label{mechanisms}
The previous field observations and measurements rise several
questions.  What are the basic dynamical mechanisms acting in the
initiation and the propagation of barchans?  What determines the
equilibrium shape of bar\-chans?  Can we predict the morphologic
relationships and the speed of dunes?  Why is barchan shape not scale
invariant?

The first part of this program i.e.  the basic dynamical mechanisms,
have been investigated in Bagnolds pioneering work \cite{B41}.  We
give here a complete overview of these mechanisms and we reformulate
the corresponding scaling laws.

\subsection{Saltation and reptation}
\subsubsection{Turbulent boundary layer}
The dune dynamics is controlled by the sand transport which is itself
driven by the wind.  For a fully developed turbulent wind over a flat
surface composed by grains of typical size $d$, the wind velocity $u$
usually increases logarithmically with height $z$ (see
figure~\ref{LogProfile}).  This can be simply understood in the
context of turbulent boundary layers theory. The standard turbulent
closure relates the air shear stress $\tau$ to the velocity gradient
$\partial_{z} u$:
\be
\label{turbulentclosure}
\tau=\rho_{air} \left(\kappa \frac{\partial u}{\partial \ln z}\right)^2
\ee
where $\rho_{air}$ is the volumic mass of air (see
table~\ref{NumericalValues}) and $\kappa \simeq 0.4$ is the Von
K\'arm\'an constant.  For a steady and uniform boundary layer, the
shear stress $\tau$ is constant and equal to $\tau_0$, the shear force
per unit area on the bed.  Besides, it turns out that the velocity
vanishes at a distance $r d$ from the sand bed.  The rescaled bed
roughness $r$ is found to be of the order of $1/30$ \cite{B41,CWG93}. This
gives a logarithmic velocity profile, as observed on the field:
\be
\label{turbulentprofile}
u(z)=\frac{u_*}{\kappa} \ln{\frac{z}{r d}},
\ee
where the shear velocity $u_*$ is by definition
$u_*=\sqrt{\tau_0/\rho_{air}}$.  We will see in the following how the
sand transport modifies this velocity profile.

\subsubsection{Motion of one grain in the wind}
A wind of sufficient strength can dislodge and entrain sand grains.
Once they have taken off the sand bed, they are progressively
accelerated by the wind, as they go up.  The trajectories are observed
to be asymmetrical, as shown on figure~\ref{LogProfile}.  The grains
are submitted to the gravity and to the fluid drag force which leads
to the standard following equation:
\be
\label{dvdt}
\frac{d \vec v}{dt}=\vec g +
\chi~\frac{\rho_{air}}{\rho_{sand}}~\frac{|\vec u-\vec v
|(\vec u-\vec v)}{d} \ee
where $\vec v$ is the grain velocity, $\rho_{sand}$ the volumic mass
and $\vec u$ the local wind velocity -- which depends on $z$.  The
drag coefficient $\chi$ depends on the shape of the grain but also on
the Reynolds numbers $Re=d~|\vec u-\vec v |/\nu$ ($\nu$ stands for the
air viscosity).  At large $Re$, in the turbulent regime, $\chi$ tends
towards a constant if the grain is sufficiently rough \cite{JS86}.  In
the viscous -- Stokes -- regime, $\chi$ decreases as $1/Re$.
\bfig[h!]  \bc \epsfxsize=\linewidth \epsfbox{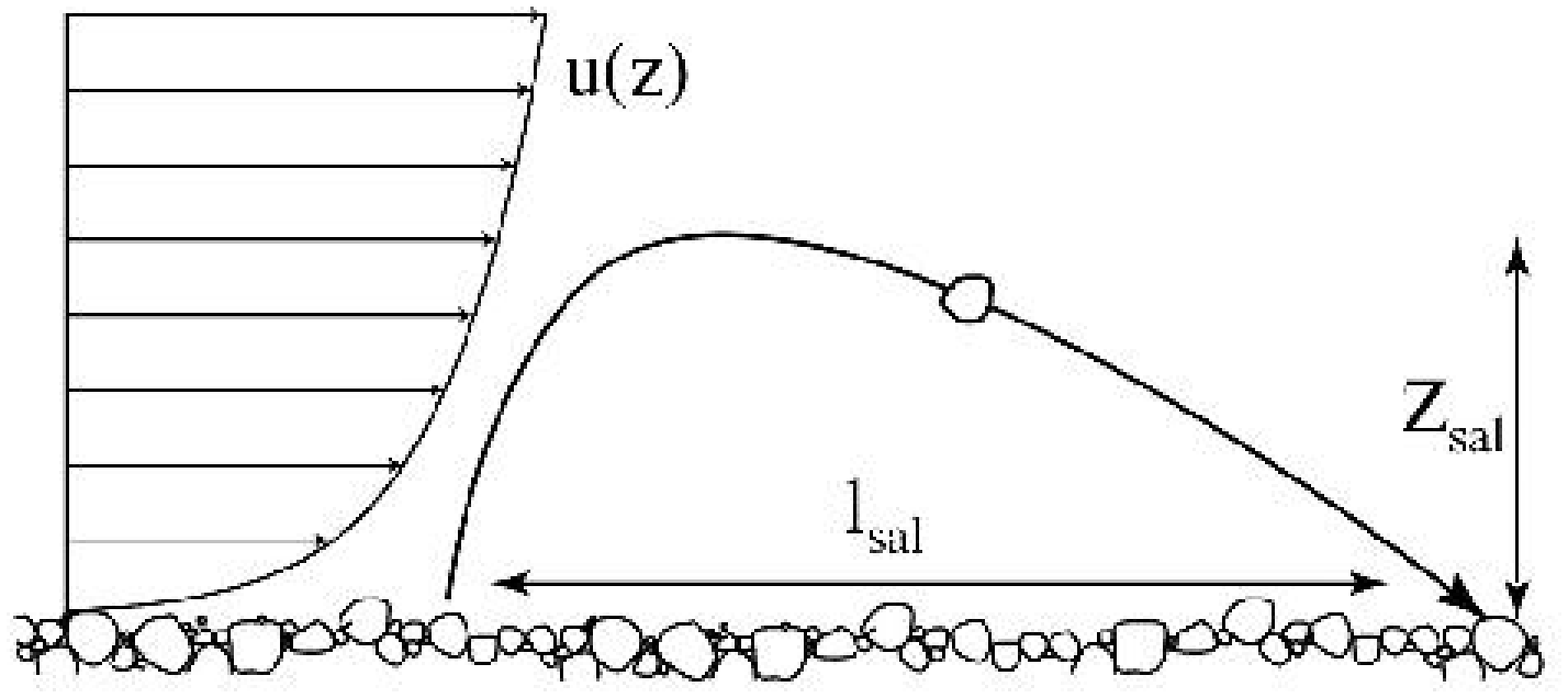}
\caption{In a fully turbulent boundary layer, the velocity profile is
logarithmic.  A grain dislodged from the bed is accelerated by the
wind as it goes higher and higher and decelerated in the descent so
that its typical trajectory is asymmetric.  Saltation hop lengths
$l_{sal}$ are found to be about $12$ to $15$ times the height of
bounce $Z_{sal}$.}
\label{LogProfile}
\ec
\efig

In the vertical direction, the motion is dominated by gravity.  It
immediately follows, as for a free parabolic flight, that the flight
time $T$ and the height of bounce $h$ depend on the launch speed $w$
as:
\be
T \propto \frac{w}{g}; \qquad Z \propto \frac{w^2}{g}.
\ee
Note that, due to the quadratic drag force, these expressions must be
corrected at strong winds \cite{NHB93}: the height of bounce $Z$ get
actually smaller than that of the free flight. These scalings are valid
whatever the launch velocity $w$ is. In the sequel we will
distinguish between the different kinds of trajectories. For example,
the indexes `sal' and `rep' will be used for saltation and reptation
grains. If no label is specified, it means that the argument applies
in general.

How does the flight time $T$ compare with the timescale $T_{drag}$
after which the grain has been accelerated by the drag force up to the
wind velocity?  Using the equation of motion (\ref{dvdt}), we get:
\be
T_{drag} \propto \frac{\rho_{sand}~d}{\chi~\rho_{air}~u},
\ee
For typical values of the parameters (table~\ref{NumericalValues}),
the saltation flight time $T_{sal}$ and the drag timescale $T_{drag}$
turn out to be of the same order of magnitude.  This means that grains
-- larger than $100 \mu m$ -- are not much sensitive to the turbulent
fluctuations of the wind but also that the horizontal velocity of the
grain is not far from the mean wind velocity along the trajectory.
The grain spends most of the flight time around the trajectory maximum
($z \simeq Z$).  As a consequence, the grains mean horizontal velocity
should be approximately equal to the wind velocity $u$ at this height.
The horizontal displacement -- the saltation length -- is of the order
of the flight time $T$ times the mean horizontal velocity:
\be
l \propto \frac{w u}{g}.
\label{Saltation_Length}
\ee

All the direct \cite{B41,NHB93} and indirect \cite{S85,JS86} measurements
as well as the models \cite{B41,O64,S91} are in agreement with this
simple approach of grains trajectories.  On the other hand, the
scalings of $Z$ and $l$ with the wind shear velocity $u_*$ and the
grain diameter $d$ are still controversial up to now.  Mainly, the
problems are the modification of the wind -- and thus $u$ -- by the
saltating grains and the mechanism by which the launch velocity $w$ is
selected.  As from now, it can be argued that the fastest grains,
which are said to be in saltation, bounce so high that the wind $u$ at
the trajectory maximum $Z_{sal}$ is almost undisturbed by the rare
grains to pass there.  Their velocity, in particular the vertical
component $w$ after a collision, should thus scale as $u_*$.  Taking
$w=u_*$, as suggested by Owen \cite{O64}, the height of bounce reads:
\be
Z_{sal} \simeq \frac{u_*^2}{g}.
\ee
The horizontal velocity $u$ (at $z=Z_{sal}$) scales with $u_*$ but
with a non dimensional prefactor ($u \simeq \xi~u_*$) reflecting
the logarithmic velocity profile:
\be
\xi=\frac{1}{\kappa}~\ln{\left(\frac{u_*^2}{rgd}\right)}.
\ee
For those high energy grains, the saltation length,
\be
l_{sal} \simeq \xi \frac{u_*^2}{g}
\ee
is $\xi$ times larger than the saltation height $Z_{sal}$
(figure~\ref{LogProfile}).  Experimentally \cite{B41,NHB93}, the ratio
$l_{sal}/Z_{sal}$ is between $12$ and $15$, which is the order of
magnitude found here for $\xi$ (table~\ref{NumericalValues}).  The
grains bouncing the highest should thus have trajectories almost
independent of their size $d$ and scaling on $u_*^2/g$ for the lengths
and on $u_*$ for the velocities.
\begin{table}[t!]
\caption{Definition and typical values of the main quantities discussed in the
text.}
\begin{center}
\label{NumericalValues}
\begin{tabular}{lll}
Dune length&$L$ &\\
Dune height&$H$ &\\
Dune width&$W$ &\\
Length of dune horns&$L_{horns}$ &\\
Dune velocity&$c$&\\
Dune profile&$h(x,y,t)$&\\
Sand flux&$\vec q(x,y,t)$&\\
Rescaled roughness & $r$ & $1/30$\\
Von K\'arm\'an constant & $\kappa$ & $0.4$\\
Shear stress & $\tau$ & \\
Air-borne shear stress & $\tau_{air}$ & \\
Grain-borne shear stress & $\tau_{sand}$ & \\
Quartz density&$\rho_{sand}$ & $ 2650~kg.m^{-3}$\\
Grain diameter&$d $ & $ 200~\mu m$\\
Shear velocity&$u_* $ & $ 0.5~m.s^{-1}$ \\
Logarithmic velocity prefactor&$\xi $ & $ 20$\\
Wind velocity&$u $ & $ 10~m.s^{-1}$\\
Saltation flight time&$T_{sal} $ & $ 50~ms$\\
Saltation hop height&$Z_{sal} $ & $ 2.5~cm$\\
Saltation hop length&$l_{sal} $ & $ 50~cm$\\
Reptation hop length&$l_{rep} $ & $ 8~mm$\\
Static friction coefficient&$\mu_s$ & $\tan(33\deg) \simeq 0.65$\\
Dynamic friction coefficient&$\mu_d $ & $ \tan(31\deg) \simeq 0.60$\\
Air density&$\rho_{fluid} $ & $ 1.2~kg.m^{-3}$\\
Air viscosity&$\nu $ & $ 1.5~10^{-5}~m^2.s^{-1}$\\
Grain Reynolds number&$Re$ & $ d u_*/\nu \simeq 120$\\
                     && $ d u  /\nu \simeq 2500$\\
Impact threshold velocity&$u_{imp} $ & $ 15~cm.s^{-1}$\\
Fluid threshold velocity&$u_{flu} $ & $ 20~cm.s^{-1}$\\
Saturated sand flux&$q_{sat} $ & $ 180~m^2 / year$\\
Drag time&$T_{drag} $ & $ 50~ms$ \\
Drag length&$l_{drag} $ & $ 9~m$\\
Reptation flux&$q_{rep}$ &\\
Vertical sand flux&$\oldphi$&\\
Number of splashed grains&$N_{eje}$&\\
\end{tabular}
\end{center}
\end{table}

\subsubsection{Collision of one grain on the sand bed}
\label{splash}
When a saltating grain collides the sand bed, it rebounds but can also
eject other grains (figure~\ref{Splash}).  In general, the rebounding
grain differs sufficiently from the ejected grains to be recognised.
However, there are low probability configurations for which the
incident grain delivers its momentum to few nearby grains but remains
trapped, even for large impact velocities.  The non dimensional
parameter which controls the rebound probability $p_{reb}$ is the
ratio of the impact velocity $v_{imp}$ to the velocity necessary to
escape from the potential trapping at the sand bed surface
\cite{QADD00}, namely $\sqrt{gd}$.  In particular $p_{reb}$ vanishes
when $v_{imp}$ becomes smaller than this escape velocity.  The rebound
probability resulting from the numerical simulation of Anderson and
Haff \cite{AH88,AH91} can be expressed as:
\be
\label{preb}
p_{reb}=p_{\infty}\left[1-\exp\left(-\frac{v_{imp}}{a
\sqrt{gd}}\right)\right] \ee
$p_{\infty} \simeq 0.95$ is the rebound probability for velocities
much larger than $a \sqrt{gd}$.  $a$ is a non-dimensional number equal
to $10$ in Anderson and Haff two-dimensional simulations
\cite{AH88,AH91}.

Experiments \cite{NHB93,RVB00} and numerical simulations
\cite{AH88,AH91} show that the rebound velocity $v_{reb}$ is a
fraction of the impact velocity $v_{imp}$: $v_{reb}=\gamma~v_{imp}$.
The restitution coefficient $\gamma$ is around $0.5$.  The rebound
angle $\theta_{reb}$ is almost independent of the impact velocity
(modulus and angle) and ranges from $35\deg$ to $50\deg$.
\bfig[h!]
\bc
\epsfxsize=\linewidth
\epsfbox{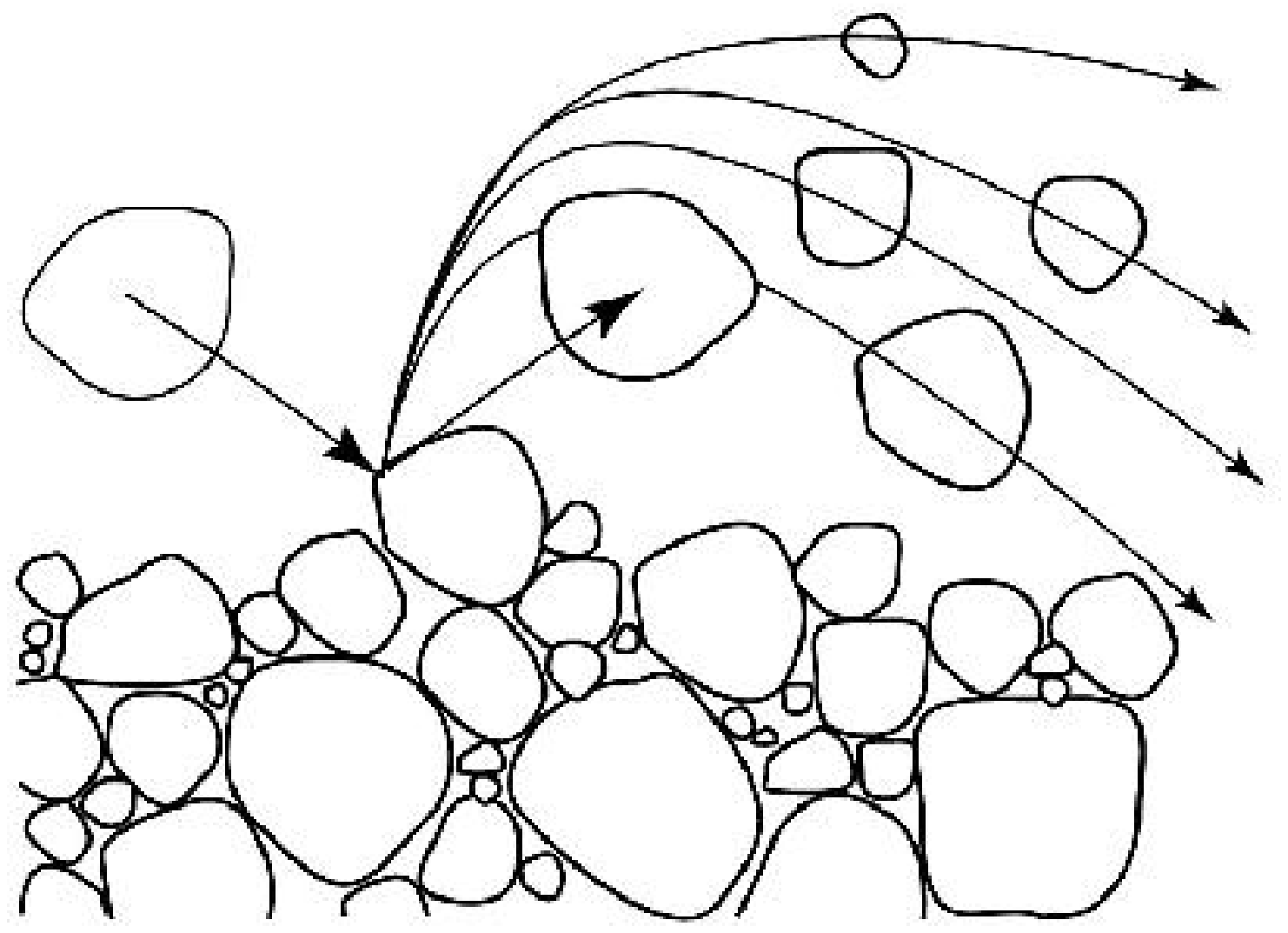}
\caption{When a saltating grain collides the sand bed, it rebounds and
splashes up other grains.}
\label{Splash}
\ec
\efig

The characteristics of the ejecta are also well documented
\cite{AH88,AH91,RVB00}.  The distribution of the ejection velocity --
both the modulus $v_{eje}$ and the angle $\theta_{eje}$ from the
horizontal -- is found to be almost independent of the impact velocity
$v_{imp}$.  The mean ejection speed is related to the escape velocity
$\sqrt{gd}$:
\be
v_{eje} \simeq a \sqrt{gd} \quad \mbox{and} \quad\theta_{eje} \simeq 70\deg
\ee
A constant fraction of the impact momentum is transferred to the
ejecta so that, on the average, the number of ejecta increases
linearly with the impact speed:
\be
\label{NEje}
N_{eje}=\frac{v_{imp}}{a \sqrt{gd}}-1\quad
\mbox{if}~v_{imp}>a \sqrt{gd} \ee
Again, the same critical velocity $a \sqrt{gd}$ appears.  It is at the
same time the critical velocity below which no grain is on the
average ejected, the mean ejection speed and the velocity below which
the rebound probability strongly decreases.

These low energy grains ejected from the sand bed move near the
surface of the sand bed: they bounce typically at a few hundred times
the grain size.  Such grains are said to be in reptation \cite{ASW91},
in contrast with saltation which relates to high energy bouncing
grains transported by the wind.  The reptation is at the origin of the
formation of ripples which, by the way, propagate and thus also takes
part in the sand transport.

\subsubsection{Saltation versus reptation}
The slowest grains, the `reptons', have properties that depend only on
the grain size: the bouncing height scales with $d$ and the grain
velocity with $\sqrt{gd}$.  On the other hand, the fastest grains, the
`saltons', have properties which depend mainly on the shear velocity
$u_*$.  Between the two, there is a continuum of trajectories of
different jump lengths which all contribute to the sand flux
\cite{AH91,RI96,RIR96}.  Experimentally, this can be observed directly
or deduced from the fact that the sand transport is a continuous
function of height.  Why then making a distinction between reptation
and saltation?

The first answer is of course that the two limit trajectories --
slowest and fastest -- of the distribution do not present the same
scaling.  Let us turn to the experiment presented in section
\ref{MinimumSize} (figures~\ref{Schema_Soufflerie} and
\ref{Film_Soufflerie}) which clearly illustrates the difference.  For
any wind velocity above the threshold of motion of the grains, the
saltation length is larger than the size of the pile (typically $1~m$
compared to $40~cm$): no saltating grain dislodged from the pile is
deposed on it.  As a consequence, the pile is only eroded.  Then it is
surprising to observe on figure~\ref{Film_Soufflerie} the formation of
horns, downwind the initial position of the pile.  This means that
they have been formed by grains deposed there which can only be grains
in reptation.  The whole figure~\ref{Film_Soufflerie} can be read in
this way: the pile is eroded due to saltation but the formation of a
crescent shape in particular the horns are due to reptation.  In this
situation, the flux of grains in saltation (visible by the overall
leak of matter) and the flux of grains in reptation (visible by the
formation of horns) are comparable.

As indicated by experiments \cite{WM91,RI96,RIR96} and numerical
simulations \cite{AH88,AH91} the sand transport is maximum at the
ground level and decreases exponentially with height.  Thus, the
contribution of reptons to the overall flux of sand should be
important.  The ratio between saltation and reptation fluxes has only
been measured directly by Bagnolds \cite{B41} using two different
traps: a rectangular trap was placed vertically for the saltons and a
second one, with a thin linear mouth, was buried in the soil to trap
the grains moving just at the surface.  In fact this hole could only
trap the slowest and largest reptons.  Bagnolds proposed to designate
this motion by surface creep.  Anderson, S\o rensen and Willetts
\cite{ASW91} gave a more precise definition of creeping grains as
grains which get rearranged by saltation impacts and which are not
affected directly by wind forces.  This category is actually useful
for binary mixtures \cite{B41}, to describe the motion of the heavy
grains submitted to a rain of light ones.  With these traps, Bagnolds
found that the creeping flux was only three to four times smaller than
the saltation flux.
\bfig[t!]  \bc \epsfxsize=\linewidth \epsfbox{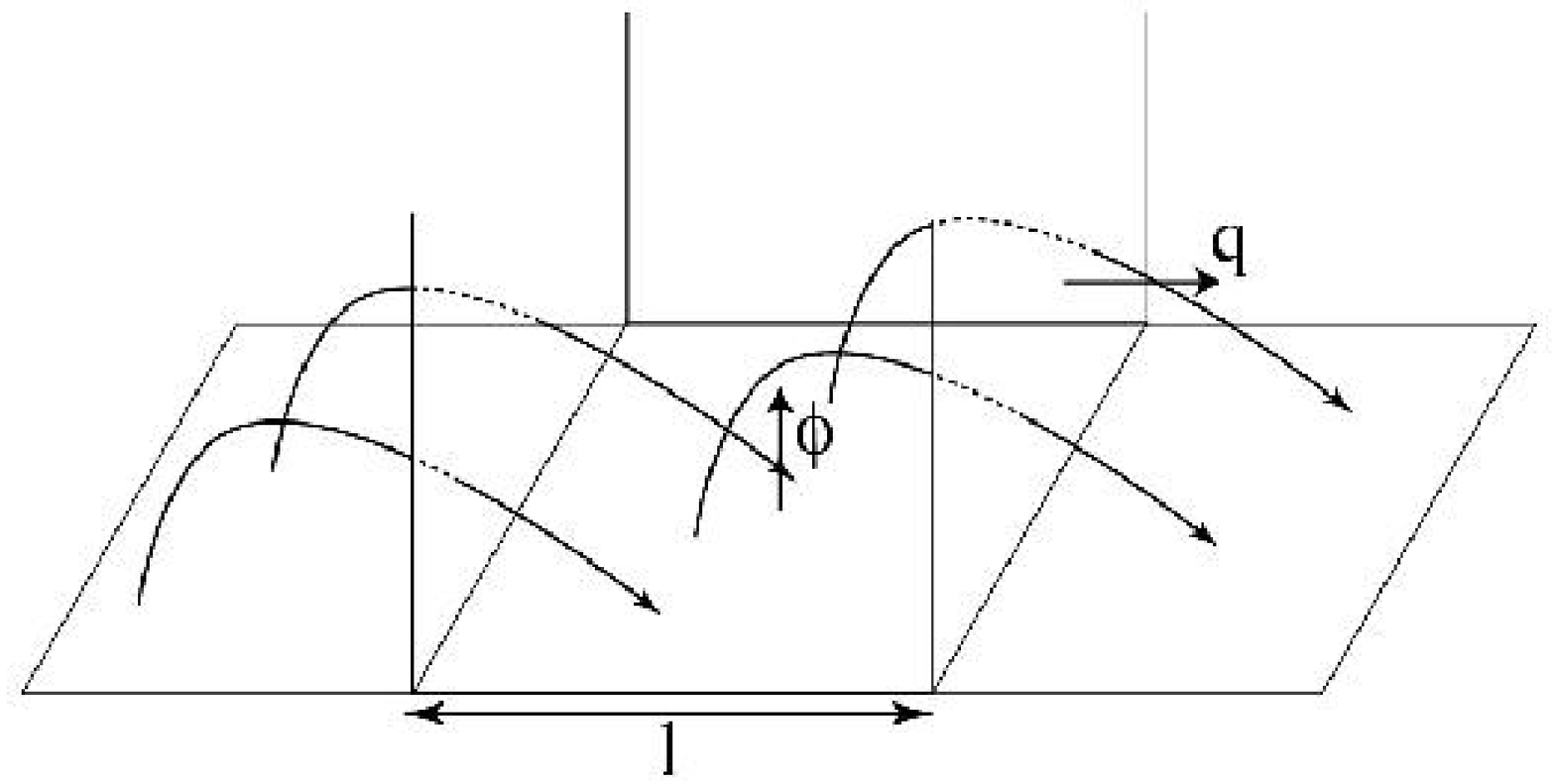}
\caption{The flux $q$ is the volume of sand which crosses a unit line
transverse to the wind per unit time.  If the typical path of a grain
has a hop length $l$, the incident flux of grains $\oldphi$, which is the
volume of grains colliding a unit area per unit time, is equal to
$q/l$.}
\label{Fluxes}
\ec
\efig

Since the grains in reptation are dragged along by saltating grains,
we can infer that the reptation flux is just proportional to the
saltation flux.  Defining the vertical flux $\oldphi$ as the volume of
particles leaving the soil per unit time and unit area, the
reptation flux $\oldphi_{rep}$ is governed by the splash of saltation
grains. In first approximation, this can be written as:
\be
\label{repsal1}
\oldphi_{rep}=N_{eje}~\oldphi_{sal}
\ee
Because $N_{eje}$ increases linearly with $u_*/\sqrt{gd}$, the number
of reptons leaving the soil per unit time is much larger than the
number of saltons.  For the sand transport, the important quantity is
the flux $q$ defined as the volume of sand crossing a unit line
transverse to the wind per unit time (figure~\ref{Fluxes}).  The
relation between the vertical flux $\oldphi$ and the integrated
horizontal flux $q$ is determined by the grains typical hop length
$l$.  For the grains in saltation, this gives:
\be
\label{qPhi}
q_{sal}=\oldphi_{sal}~l_{sal}
\ee
Then the relation (\ref{repsal1}) between fluxes of saltation and
reptation can be expressed in terms of $q$:
\be
\label{repsal2}
q_{rep}=\frac{l_{rep}~N_{eje}}{l_{sal}}~q_{sal}
\ee
Since $l_{rep}/l_{sal}$ decreases as $gd/u_*^2$, this means that the
contribution of reptons to the overall flux of sand decreases with the
wind velocity.  This is consistent with the disappearance of ripples
at large wind.  One should thus be very careful in identifying the
species most important for the sand flux: the result is opposite for
the horizontal flux $q$ and the vertical flux $\oldphi$.  Finally, the
same question can be risen: if the two fluxes are proportional one
with the other, why making a distinction between the two?

Another important difference appears when the sand bed presents a
slope, in particular if the steepest slope is perpendicular to the
wind.  Howard \cite{H77} have observed that the normal to ripples are
no more parallel to the wind on sloping surfaces but are deflected
downslope by as much as $35\deg$.  This clearly shows that the gravity
force has an influence on the direction of motion of reptons.
Moreover, Howard \cite{H77} found a good agreement with a simple
calculation considering that the ejected grains are submitted to an
effective drag along the wind direction, just sufficient to escape
from potential trapping.  Most of their flight, saltons are dragged
along the wind direction but their rebound direction could be
influenced by the surface slope.

More recently, Hardisty and Whitehouse \cite{HW88} have shown that the
bedslope along the direction of motion modifies the threshold of
motion (typically $15\%$ for a $10\deg$ slope) but also the flux
itself.  They found that the flux was multiplied by $10$ for a $10
\deg$ downslope and divided by $6$ for a $10 \deg$ upslope.  These
very striking results were not confirmed by Rasmussen and Iversen
\cite{RI96,RIR96,IR94} who have found a much weaker dependence of the
saturated saltation flux with the longitudinal slope.

There still misses experimental studies investigating the dependence
of the reptation flux with the slope -- in particular a lateral slope.
Still, it can be argued that gravity is important in the ejection
process.  The number of ejecta should be smaller for a positive slope
along the wind direction than for a negative slope and reptons should
be deflected downslope by gravity.  Then, the total flux is not
aligned along the direction of wind but has a component along the
steepest slope.  To the first order this downslope flux is simply
proportional to the gradient $\nab h$ of the quantity transported (the
local height $h$).  It thus leads to diffusive effects which and tend
to smooth the dune.

\subsection{Entrainment by the fluid and by impacts}
\subsubsection{Fluid threshold}
\bfig[h!]
\bc
\epsfxsize=\linewidth
\epsfbox{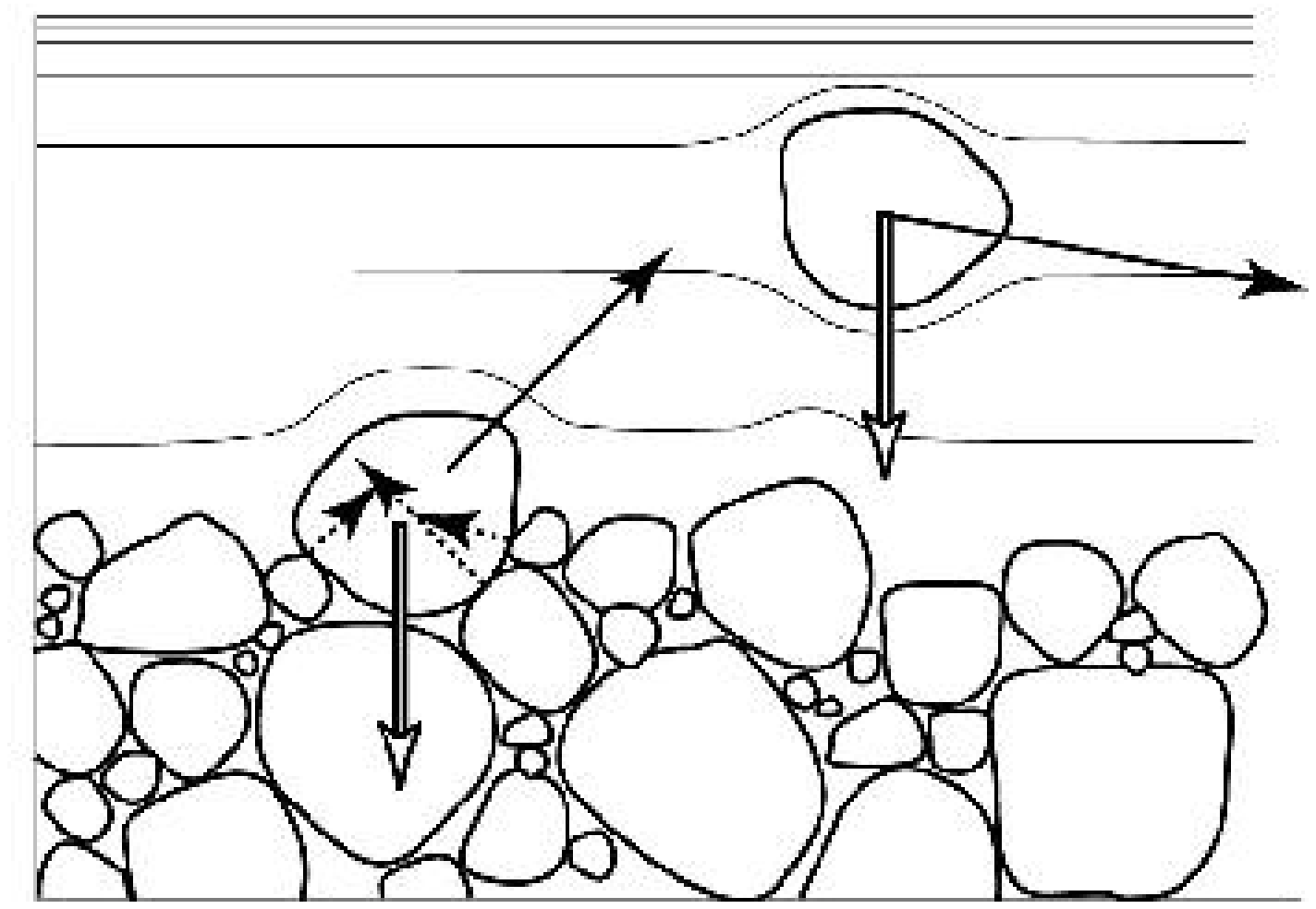}
\caption{Wind is accelerated over grains at the surface of the sand bed.
To lift up a grain, the low pressure created by the wind flow has to be larger
than gravity. There is thus a threshold velocity below which the wind
can not erode the sand bed.}
\label{Threshold}
\ec
\efig

It is somewhat striking that grains can be dislodged by the wind since
they fall down again after having taken off.  This is due to the
asymmetry of the flow around a static grain at the surface of the sand
bed, as demonstrated by figure~\ref{Threshold}.  Since $\tau$ is by
definition the horizontal force per unit area, the drag force acting on
a grain at the surface of the sand bed scales as $\tau_0 d^{2}$.  A
grain will escape from trapping only if this force is larger than
gravity.  As a consequence, there exists a threshold shear velocity
$u_{flu}$ below which a sand bed at rest cannot be eroded by the
wind:
\be
u_{flu} = \zeta_{flu} \sqrt{\frac{\rho_{sand}-\rho_{air}}{\rho_{air}}~gd}.
\ee
Experimentally, the rescaled threshold velocity $\zeta_{flu}$ is found
to be around $0.1$ (see table~\ref{NumericalValues}).

Above $u_{flu}$ grains spontaneously start rolling at the surface of
the sand bed \cite{WMR91}.  During their motion, some grains can take
off the bed, due to the bumps beneath them or to the aerodynamic lift
force.  As soon as they have left the bed, the flow around them become
again symmetric (figure~\ref{Threshold}) so that they start being
accelerated downward by gravity. They collide the bed, rebound and
eject other grains.  The later are accelerated by the wind, splash on
the sand bed and so on, until saturation be reached (see
part~\ref{saturationpara}).

\subsubsection{Impact threshold}
Suppose that there are already some grains in saltation.  If the wind
velocity decreases, the grains velocity also decreases.  So does the
sand flux.  The minimum wind velocity which can sustain the sand
transport is that for which the grains impact velocity is no more
sufficient to eject other grains (see section~\ref{splash}).  This
happens for $v_{imp} \simeq a \sqrt{gd}$ and thus for a wind velocity
$u_*$ of the order of a few times $\sqrt{gd}$:
\be
u_{imp} = \zeta_{imp} \sqrt{gd}.
\ee
Experimentally, the rescaled threshold velocity $\zeta_{imp}$ is found
to be around $3.5$ (see table~\ref{NumericalValues}).
\begin{table}[t!]
\caption{Typical values of the quantities defined and discussed in the
text. The aeolian sand transport is compared to that in water.}
\begin{center}
\label{ComparisonAirWater}
\begin{tabular}{lll}
Fluid & Air & Water\\
$\rho_{fluid}$ & $1.2~kg.m^{-3}$& $1000~kg.m^{-3}$\\
$\nu$ & $1.5~10^{-5}~m^2.s^{-1}$ & $10^{-6}~m^2.s^{-1}$\\
$Re_*=d u_*/\nu$ & $120$ & $8$\\
$Re=d u/\nu$ & $2500$ & $170$ \\
$u_{imp}$ & $15~cm.s^{-1}$ & $15~cm.s^{-1}$\\
$u_{flu}$ & $20~cm.s^{-1}$ & $0.7~cm.s^{-1}$\\
$q_{sat}$ & $180~m^2 / year$ & $ 17~m^2/hour$\\
$T_{drag}$ & $50~ms$ & $ 0.1~ms$ \\
$l_{drag}$ & $9~m$ & $1~cm$\\
\end{tabular}
\end{center}
\end{table}

\subsubsection{Entrainment by the fluid and by impacts}
In the case of aeolian sand transport, we see on
table~\ref{ComparisonAirWater} that the `impact' threshold velocity is
smaller than the `fluid' threshold velocity.  This means that the
transition from the sand bed at rest to the saturated sand transport
presents a hysteresis.  Starting from a light wind ($u_*$ smaller than
both $u_{imp}$ and $u_{flu}$), the sand flux is null.
Then, if the wind velocity increases above the fluid threshold
$u_{flu}$, the grains start being entrained directly by the wind
and a saturated sand transport establishes.  Now, if the wind velocity
decreases, the sand flux vanishes at $u_{imp}$ only.  This is
characteristic of a subcritical transition.  The consequences of this
hysteretic behaviour has not be investigated, so far.

>From this point of view, the sand transport behaves very differently
in liquids.  It can be seen on table~\ref{ComparisonAirWater} that the
fluid threshold velocity in water is much smaller than the impact
threshold.  This means that the hysteresis disappears in water --
as in liquids in general.  It also indicates that the sand transport
under water is dominated by direct fluid entrainment, due to the
favourable density ratio.  By comparison, the splash process is
inefficient and there is, so to say, no saltation -- and therefore no
reptation -- under water.

In order to distinguish the direct entrainment of grains by the fluid
from saltation and reptation, we propose the name `tractation' for
this motion.  This word is derived from the latin verb {\it trahere}
which means to drag \footnote{{\it Saxa ingentia fluctus trahunt} (De
bello Jugurthino, Sallustius), i.e.  huge stones are dragged by the
flow.}.  `Tractons' are thus the predominant species under water.
Tractation is very similar to avalanches except that the driving
mechanism is the fluid drag instead of the gravity.  Then, the same
kind of description could in principle be adapted (see
section~\ref{Avalanches}).

\subsection{Saturated flux and saturation length}
\label{saturationpara}
\subsubsection{Negative feedback on the wind}
The only limit to the erosion of a sand patch is the saturation of the
sand flux.  It occurs because of the negative feedback of the
transported sand on the wind strength.  The point is that the same
momentum flux is used to maintain the turbulent boundary layer and to
speed up the entrained grains.  For a given shear velocity, there is
thus a maximum flux of sand, called the saturated flux $q_{sat}$,
which can be transported by the wind.

Since they are accelerated by the wind, the grains, at a height $z$,
have a smaller velocity $u_{\uparrow}$ when they go up than when they
come down again ($u_{\downarrow}$).  The volume of sand which collides
a unit area of the sand bed per unit time is, by definition, the flux
$\oldphi$.  As shown on figure ~\ref{Fluxes}, it is related to the
horizontal flux $q$ integrated along the vertical and to the hop
length $l$: $\oldphi=q/l$.  The momentum transferred to this volume of
sand is by definition the sand-borne shear stress $\tau_{sand}$ and is
equal to the mass flux $\rho_{sand}~\oldphi$ times the velocity
difference $u_{\uparrow}-u_{\downarrow}$:
\be
\label{tausand}
\tau_{sand}=\rho_{sand}~\oldphi~(u_{\uparrow}-u_{\downarrow})
\ee
The remaining part of the total shear stress, the air-borne shear
stress $\tau_{air}=\tau_0-\tau_{sand}$, accelerates the flow itself.
Assuming that the turbulent boundary layer is still at equilibrium,
the standard turbulent closure (\ref{turbulentclosure}) leads a
modified velocity profile given by:
\be
\frac{\dr u}{\dr \ln
z}=\frac{1}{\kappa}\sqrt{u_*^2-\frac{\rho_{sand}}{\rho_{air}} \oldphi
(u_{\uparrow}-u_{\downarrow})}
\ee
with the same boundary condition than previously, $u=0$ at the height
$z=r d$.
\bfig[t!]  \bc \epsfxsize=\linewidth \epsfbox{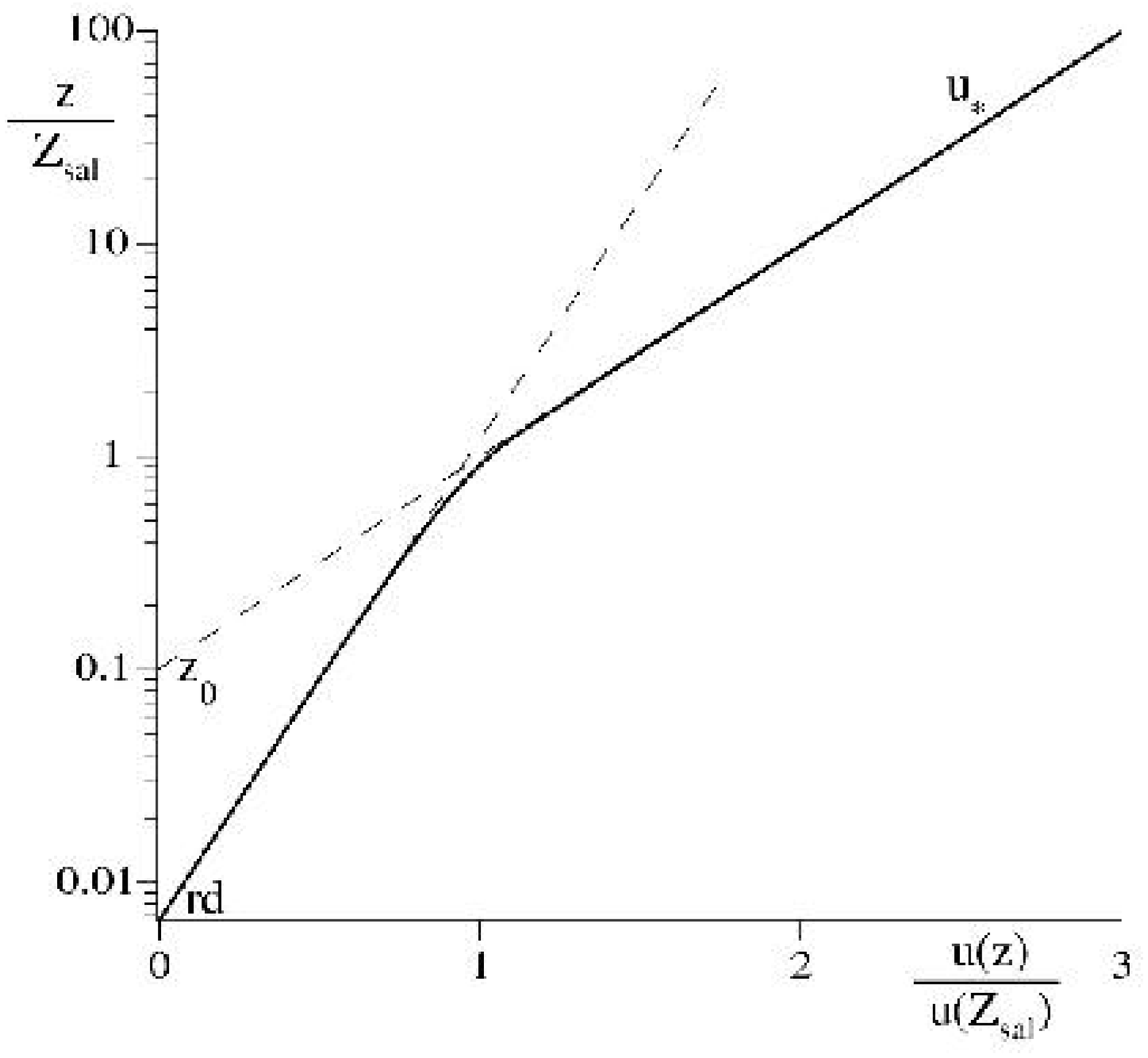}
\caption{Velocity profile modified by the sand transport, as obtained
theoretically by Owen \cite{O64} and Raupach \cite{R91} and in
numerical simulations by Anderson and Haff \cite{AH88,AH91}.  It is a
piecewise logarithmic profile.  Inside the saltation layer
($z/Z_{sal}<1$), the shear velocity is decreased but the roughness $rd$
is that of the bed.  Outside ($z>Z_{sal}<1$), the shear velocity is
$u_*$ but the apparent roughness $z_0$ has increased, as if the soil
was higher.}
\label{Curtain}
\ec
\efig

Above the saltation layer, there is no grain and thus no grain-borne
shear stress: $\tau_{air}=\tau_0$.  An undisturbed turbulent boundary
layer is thus recovered, but with an increased apparent roughness $z_0$
(figure~\ref{Curtain}).  Everything looks as if the soil has been
risen at a height $z_0-rd$.  Just above the soil, inside the
reptation layer, the air-borne shear stress $\tau_{air}$ is strongly
reduced, and is much smaller than $\tau_0$.  As a consequence, the
roughness is similar ($u$ vanishes at $rd$) but the apparent shear
velocity is smaller than $u_*$ (figure~\ref{Curtain}).
Experimentally, the velocity profile is found to be logarithmic, with
an increased apparent roughness $z_0$, but the lowest part of the
profile is too thin to be measured \cite{B41,NHB93,RI96,RIR96,IR94}.
Therefore, the numerical and theoretical findings for the velocity
profile inside the saltation layer should be checked experimentally.

\subsubsection{Equilibrium transport}
In the previous sections, only the consensual properties of sand
transport have been reviewed.  We now come -- in this subsection -- to
the controversial points.

It is clear that the hop height depends on the launch velocity $w$ and
scales as $w^2/g$.  The hop length depends on $w$ and on the mean
horizontal velocity $u$, and scales as $w u/g$.  Now, what are the
mean velocity components $u$ and $w$ when the equilibrium is achieved?
Nalpanis {\it et al.} \cite{NHB93} have found that the mean vertical
launch velocity $w$ is about $2 u_*$ and that both $h$ and $l$ scale
on $u_*^2/g$.  On the other hand, Rasmussen and Iversen
\cite{RI96,RIR96,IR94} have reported measurements of the horizontal
grain speed showing almost no dependence of $u$ on the shear velocity
$u_*$.  The scaling found by Nalpanis \& al.  \cite{NHB93} is that of
the saltons while the scaling found by Rasmussen and Iversen
\cite{RI96,RIR96,IR94} corresponds to the reptons.  A third scaling
has been obtained by Jensen and Sorensen \cite{S85,JS86} who have used
a model to extract the same quantities from experimental measurements
of the vertical variation of the transport rate.

Another -- indirect -- measurement has also been conducted by
Rasmussen and Iversen \cite{RI96,RIR96,IR94}: they have studied
systematically the dependence of the apparent roughness $z_0$ with the
wind velocity $u_*$ and the grain size $d$.  From
figure~\ref{Curtain}, it can be inferred that $z_0$ is related to the
height of bounce $Z$ \cite{R91} and to the mean horizontal velocity
$u$ ($\simeq u(Z)$) by $z_0=Z~\exp(-u(Z)/u_*)$.  If $w$ and $u$ scale
as $u_*$, $z_0$ should scale as $u_*^2/g$.  If $w$ and $u$ are
independent of $u_*$, $z_0$ should increase with $u_*$ and saturate at
a value independent of $u_*$.  The experimental measurements can be
rescaled to give approximately:
\be
\label{zzero}
z_0 \simeq r \left(\frac{u_*^2}{g}+(1-\zeta_{imp}) d\right) \ee
The apparent roughness $z_0$ tends to the soil roughness $r d$ at the
impact threshold and above, increases and scales asymptotically as
$u_*^2/g$.  According to the previous analysis, this means that the
feedback of the sand transport on the wind is dominated by high energy
grains.

As a conclusion, the question of the trajectory mean properties is
not completely solved, perhaps because it is ill posed.  It is clear
that there is a distribution of characteristics ranging from the low
energy reptons to the high energy saltons.  Then, the scaling laws
obviously depend on the way quantities are averaged.  For instance,
the result will be different if an average is weighted by the vertical
flux, by the grain density, by the horizontal flux or even by the
energy flux.  Also the result will depend on the height of the lowest
trajectory taken into account in the statistics, for instance when a
camera is used.

\subsubsection{Saturated flux}
The saturation of the sand flux can be understood without entering in
the details of the mechanisms.  The first idea, proposed by Bagnold
\cite{B41}, was that equilibrium is reached when the sand-borne shear
stress has taken a given part of the overall shear stress:
$\tau_{sand} \propto \tau_{0} = \rho_{air} u_*^2$.  Replacing $\oldphi$
by $q/l$ in equation (\ref{tausand}), we get:
\be
q_{sat} \propto \frac{\rho_{air}}{\rho_{sand}} \frac{u_*^2~
l}{u_{\uparrow}-u_{\downarrow}}
\ee
The hop length $l$ scales as $uw/g$.  The velocity difference between
the rise and the descent $u_{\uparrow}-u_{\downarrow}$ should be a
fraction of the grain horizontal velocity $u$. This gives
\bea
\label{qsat1}
q_{sat} &\propto \frac{\rho_{air}}{\rho_{sand}} \frac{w u_*^2}{g}\quad
&\mbox{if}~u_*>u_{imp}\cr
q_{sat} &= 0\quad&\mbox{if}~u_*<u_{imp}
\eea

Owen \cite {O64} have introduced a refined argument: the saturation is
reached when the wind shear velocity inside the saltation layer has
decreased to its threshold value.  In Owen's article, this threshold
was $u_{flu}$, meaning that the erosion is due -- and limited -- by
the direct aerodynamic entrainment.  It seems more reasonable to use
the impact threshold $u_{imp}$.  Since
$\tau_{sand}=\tau_0-\tau_{air}=\rho_{air} u_*^2-\zeta_{imp} gd$, this
gives:
\be
\label{qsat2}
q_{sat} \propto \frac{\rho_{air}}{\rho_{sand}} w
\left(\frac{u_*^2}{g}-\zeta_{imp} d\right)
\ee
More detailed formulas have been proposed for this relation
\cite{LL78,S91,SKH01} which, as this one, essentially smooth the
Bagnolds relation around the threshold velocity $u_{imp}$.

Both expressions (\ref{qsat1}) and (\ref{qsat2}) depend on the typical
launch velocity $w$ which, as seen in the previous section, remains
problematic.  Nevertheless, we can conclude on the scaling of the
saturated flux.  First, the flux $q_{sat}$ associated to high energy
saltons ($w \propto u_*$) increases more rapidly than the flux
associated to low energy reptons ($w \propto \sqrt{gd}$).  Second, to
maintain a possible saturated reptation flux, a saltation flux is
needed, given by equation (\ref{repsal2}).  For large shear velocity,
the non dimensional factor $l_{sal}/(l_{rep}~N_{eje})$ relating
$q_{rep}$ to $q_{sal}$ scales as $u_*/\sqrt{gd}$.  So, to maintain a
flux saturated by reptons, a flux of saltons scaling as
$\rho_{air}/\rho_{sand} u_*^3/g$ is needed.  In any case, this is the
scaling expected at large shear velocity even if reptons can be
predominant just above the threshold.  The last -- and the best --
argument is that $q_{sat} \propto u_*^3$ is the scaling measured
experimentally \cite{B41,LL78,WM91,RM91}, with a prefactor of order unity.

The expression (\ref{qsat2}) predicts that the saturated flux should
slightly decrease for an increasing grain diameter $d$, due to the
threshold effect.  From experimental measurements, Bagnolds \cite{B41}
has reported an increase of $q_{sat}$ with $d$: for grains ranging
from $100~\mu m$ to $1~mm$, the prefactor in equation (\ref{qsat1}) is
found to be $\sqrt{d/D}$ with a new length $D \simeq 150~\mu m$.  This
is very striking since $D$ should neither depend on $u_*$ nor on $d$.
Rasmussen and Iversen \cite{RI96,RIR96,IR94} have also an extra length
in the scaling of the apparent roughness $z_0$.  Introducing the same
length $D \simeq 150~\mu m$, the fit of their data by
\be
z_0 \simeq r \sqrt{D/d} \left(\frac{u_*^2}{g}+(1-\zeta_{imp}) d \right)
\ee
is much better than with expression (\ref{zzero}).  And again, $D$
should neither depend on $d$ nor on $u_*$.  This new parameter $D$
have to be related to the fluid entrainment (with a Reynolds number
dependence) or to the collision process (with a dependence on the
elastic properties of the material).  This implies the existence of
new mechanisms, not investigated so far.

\subsubsection{Space and time lag}
When the wind arrives at the edge of a sand sheet, it dislodges and
carries away sand grains.  The later fall down again and eject other
grains so that the transport rate increases.  In turn, these grains in
saltation reacts on the turbulent boundary layer to decrease the
velocity.  There is thus a lag between the edge of the sand sheet and
the achievement of the sand flux saturation.  This was first reported
by Bagnolds \cite{B41} who observed experimentally that the transport
rate of sand nonetheless increases towards its saturated value but
overshoots and exhibits a damped oscillation in space.  Other
experimental papers report space lags but none have recovered the
oscillation.  The existence of a lag was recovered in the numerical
simulation by Anderson \& Haff \cite{AH88,AH91}.  Several authors have
observed that the space lag before saturation -- the saturation length
-- is almost independent of the shear velocity $u_*$, contrarily to
the saltation length.  Note that in cellular automaton models
\cite{NO93,W95,NY98}, there is also a lag but due to transport non
locality i.e.  to the fact that grains moves by jumps of one saltation
length $l_{sal}$.
\bfig[h!]  \bc \epsfxsize=\linewidth \epsfbox{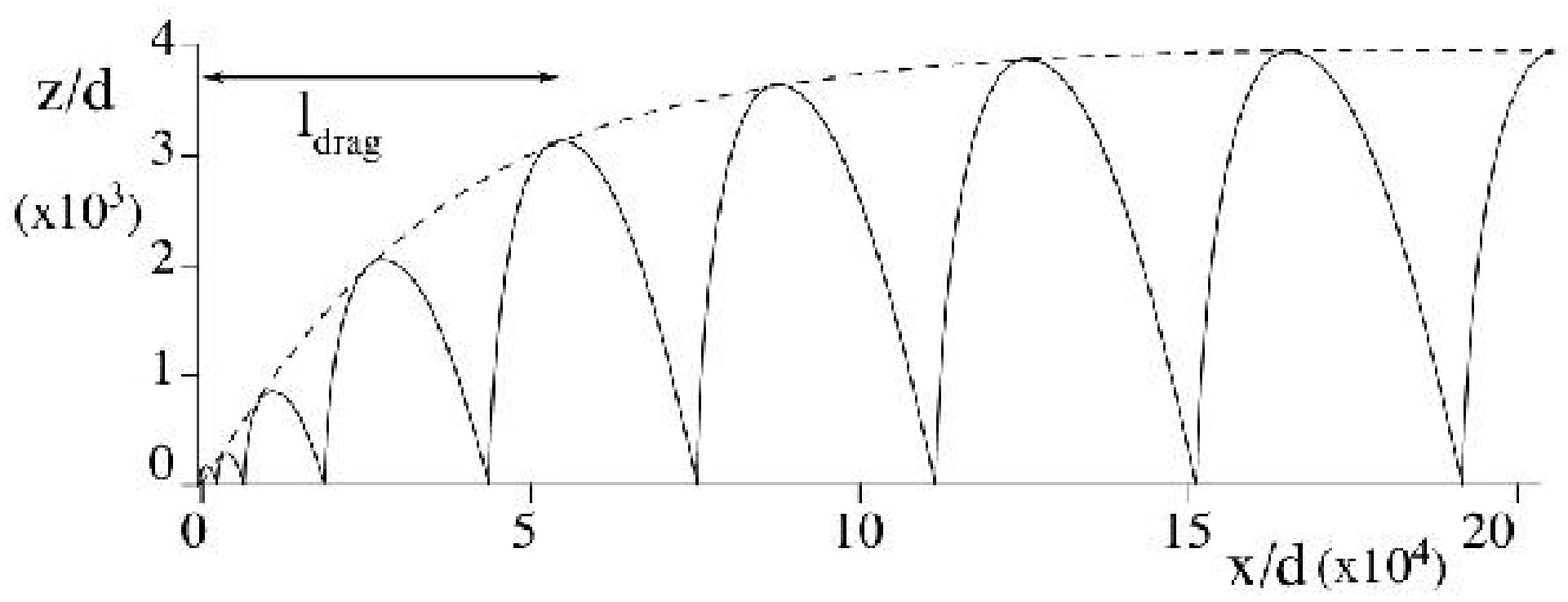}
\caption{Sketch of the promotion of a repton to saltation.  Once
ejected the grain reaches the wind velocity after a typical length
$l_{drag}$.  The equilibrium sand transport is achieved when the wind
speed has so decreased that the number of reptons promoted to
saltation is just equal to the number of saltons absorbed by the sand
bed at their impact.}
\label{Acceleration}
\ec
\efig

The physical mechanisms which lead to the sand transport equilibrium
have not to be specified to derive the saturated flux.  It is the
interest of Bagnolds and Owen arguments but also their limits.  But to
understand the origin of the saturation length, these mechanisms have
to be made explicit.  Owen \cite {O64} have proposed a first idea: the
flux saturates when the wind does not directly entrain grains any more
i.e.  when the wind shear velocity inside the saltation layer has
decreased to its threshold value $u_{flu}$.  It is probably the
case for the sand transport in water but not in air, in which the
direct aerodynamic entrainment is much less efficient than the
entrainment by impacts.

Anderson \& Haff \cite{AH88,AH91} characterise the production of new
grains in saltation by the mean replacement rate $\eta$ which is the
mean number of grains dislodged by an impacting grain (except itself,
which rebounds).  The flux saturates when the wind and thus the grains
velocity becomes so small that $\eta$ vanishes.  The erosion rate
which is the difference between the flux of grains taking off the bed
and the flux of grains impacting on it, is given by the spatial
derivative of the sand flux $\ddx q_{sal}$.  It is equal to the incident
flux of grain $\oldphi_{sal}=q_{sal}/l_{sal}$ times the mean replacement rate
$\eta$:
\be
\label{etasurl}
\ddx q_{sal} = \eta \frac{q_{sal}}{l_{sal}}
\ee
The simplest possibility is to imagine, as proposed by S\o rensen
\cite{S91}, that the ejected grains are of the same species than the
impacting ones.  $\eta$ is then simply the number of ejected grains
$N_{eje}$ described by the equation (\ref{NEje}).  The saturation is
reached when $N_{eje}$ vanishes.  This appends when all these
saltons have a velocity of the order of $a \sqrt{gd}$ -- independent
of $u_*$ -- i.e.  when they have become reptons
(figure~\ref{Curtain}).  Because each grain gives $1+N_{eje}$ grains
after a collision, the flux first increases exponentially with a
typical lengthscale $l_{sat}=l_{eje}$ equal to the saltation length
$l_{sal}$ divided by the typical replacement rate $\eta$:
\be
l_{eje} \simeq a d~\frac{u_*}{\sqrt{gd}}.
\ee

This equilibrium situation with a uniform reptation layer is far too
simplified.  If a grain jumps just above this layer, it is
progressively accelerated and becomes a salton, as seen on
figure~\ref{Acceleration}: the uniform reptation layer is unstable.
Second, low energy grains have a large probability $1-p_{reb}$ to be
absorbed by the sand bed when they collide it (eq.~\ref{preb}).  At
equilibrium, this should be balanced by a production of reptons.
This suggests a slightly different picture in which saltons and
reptons coexist pacifically.  The saltons produce reptons when
they collide the sand bed.  reptons are promoted to the rank of
saltons, once accelerated by the wind (figure~\ref{Acceleration}).
This is probably the situation reached in Anderson \& Haff numerical
simulation \cite{AH88,AH91}.  The production of reptons involves the
typical length scale $l_{eje}$ defined above.  The acceleration of
reptons to the velocity of saltons takes a length $l_{drag}$
scaling on the grain size and on the sand to fluid density ratio:
\be
l_{drag} \simeq \xi~\frac{\rho_{sand}}{\rho_{air}}~d
\ee

As time -- and space -- goes by, the reptation and saltation flux
increase and the wind strength decreases.  Equilibrium is reached when
the number of reptons promoted to saltation just balances the small
absorption of grains in collisions.  Since the process involves two
species and thus two lengthscales $l_{eje}$ and $l_{drag}$, the
saturation length $l_{sat}$ should be the largest of two.  It should also be
larger than the saltation length $l_{sat}$.  For typical values of the
shear velocity, the promotion of reptons to saltons is the
limiting mechanism so that $l_{sat}$ is given by the inertial length
$l_{drag}$.

It is interesting to note that there has been only one previous
derivation of the saturation length \cite{SKH01} which uses a
completely different argument.  Owen's criterion, predicts that the
air-borne shear stress should have decreased to its threshold value at
equilibrium.  Simultaneously, the mean replacement rate $\eta$ should
vanish.  On this basis, Sauermann {\it et al.} \cite{SKH01} proposed the
empirical scaling law $\eta \propto (u_*/u_{flu})^2-1$, which, once
reinjected in equation~(\ref{etasurl}), leads to a saturation length
$l_{sat} \simeq \xi~\rho_{sand}/\rho_{air}~d$, as found here.

Except the logarithmic dependence hidden in $\xi$ the saturation
length is independent of the shear velocity $u_{*}$.  It scales as the
grain size weighted by the sand to air density ratio and it is of the
order of $10~m$ (table~\ref{NumericalValues}), as found experimentally
\cite{B41}.

\subsection{Coupling between wind and shape}
\subsubsection{Erosion rate, minimum size and barchan speed}
In the previous section, we have seen how a sand bed is eroded by the
wind.  Now, where does the erosion takes place on a barchan and at
which rate?  To answer this question, let us introduce the local
height $h(x,y,t)$ of the dune and the volumic sand flux $\vec
q(x,y,t)$.  The conservation of matter reads:
\be
\label{conservationmatter}
\ddt h+\nab . \vec q = 0
\ee
If locally the sand flux is larger than the saturated flux then the
sand flux decreases spatially and sand is deposed.  If on the contrary
the local sand flux is smaller than the saturated flux, the sand flux
increases spatially and the sand bed is eroded.
\bfig[h!]  \bc \epsfxsize=\linewidth \epsfbox{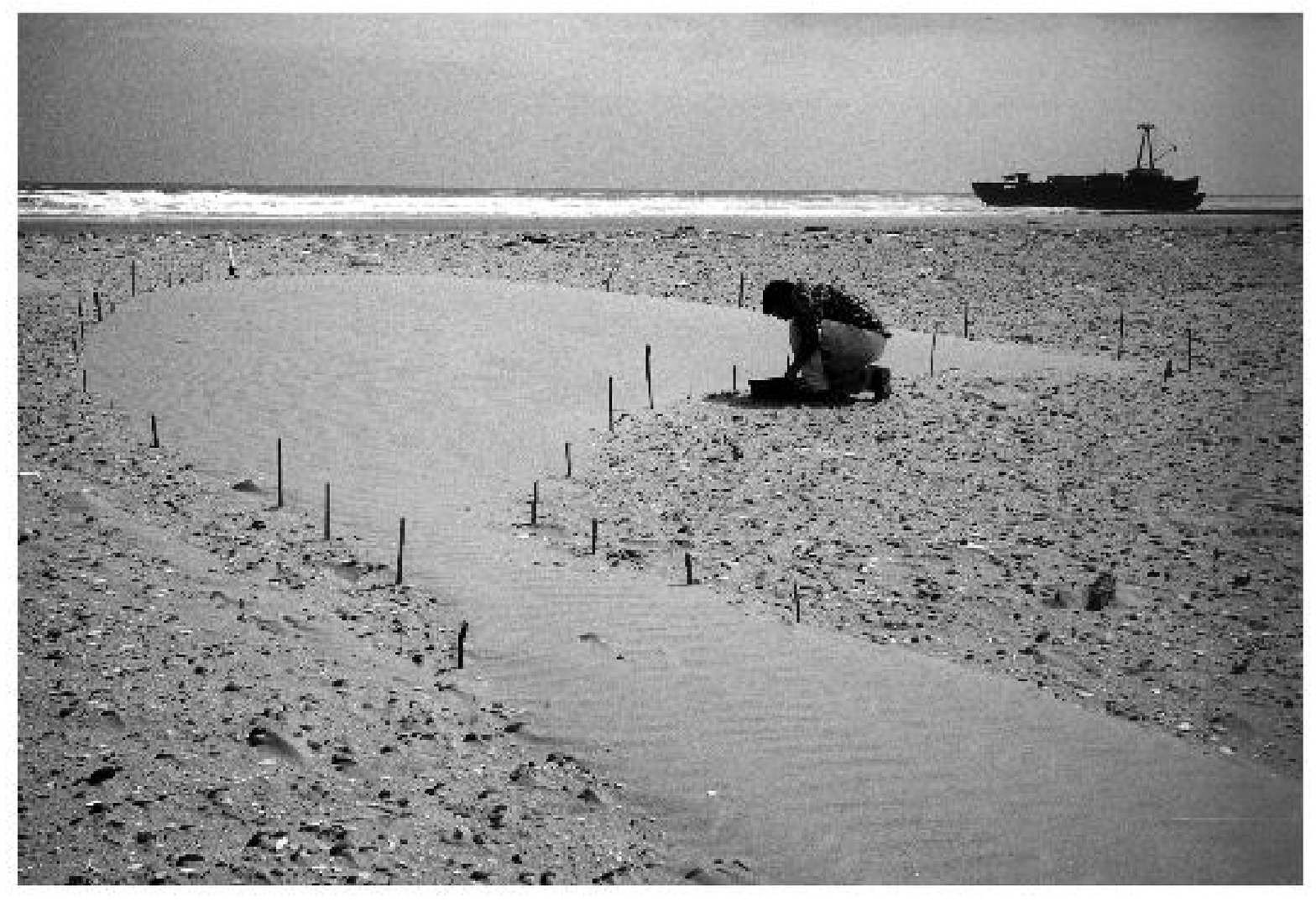}
\caption{A sand patch (dome) on Tarfaya beach (southern Morocco)
upwind the barchan field.  The precursors to barchans directly appear
with a length and a width of the order of $10~m$.}
\label{TarfayaBeach}
\ec
\efig

What appends if a dune is smaller than the saturation length
$l_{sat}$?  Because it is not saturated, the flux will increase
continuously over the whole length of the dune.  The dune will thus be
eroded everywhere and will disappear.  This explains very simply the
experimental observation reported here (figures~\ref{Schema_Soufflerie}
and \ref{Film_Soufflerie}) that a small sandpile blown by the wind
disappears.  The order of magnitude of $l_{sat}$ is also the length
(and the width) with which the sand patches which will give barchans
nucleate (figure~\ref{TarfayaBeach}).  The introduction of
the saturation length is thus very important because it explains the
existence of the minimum length of the dune.  We will see in the
subsection devoted to the wind around the dune that no other
lengthscale appears in the problem so that it could be {\it the}
relevant lengthscale in the problem.

To get a better view of the way a barchan is eroded, we can assume
that it propagates with a constant shape and speed.  Then $h$ and $q$
depend only on the variables $x-ct$ and $y$ so that the erosion rate
$\ddt h$ is equal to $- c \ddx h$.  It means that the dune is eroded
in the places where the slope along the wind direction is positive and
the sand is deposed when this slope is on the contrary positive.  When
the brink and the crest coincide (for instance the large dune on
figure~\ref{Q=vL}), the horns and the slip face are the only places
where the sand accumulates.  For a dune presenting a small slip face
(as the small dune on figure~\ref{Q=vL}) there is a third region of
accretion around the brink.

Let us consider a two-dimensional dune, invariant along the transverse
direction $y$ direction.  Under the hypothesis of constant shape and
speed, the conservation of matter (\ref{conservationmatter}) becomes
$\ddx (q_x-c h)=0$ which immediately integrates into:
\be
q(x) = q_0 + c h(x)
\label{q=cH}
\ee
where $q_0$ is the sand supply i.e.  the sand flux on the firm soil
behind the dune.  If the dune if sufficiently long, the sand flux
$q(x)$ is saturated at the crest of the dune.  Then the propagation
speed immediately derives from the equation~(\ref{q=cH}):
\be
c \simeq \frac{q_{sat}-q_0}{H}
\ee

Now, the three dimensional problem reduces to two dimensions if the
conservation of matter (\ref{conservationmatter}) is integrated along
the transverse direction $y$.  Equation~(\ref{q=cH}) thus holds if $q$
and $h$ are replaced by their average across the dune width.  We make
the reasonable assumption that the average height along a cross
section of the dune scales on the overall height $H$.  Then, the
barchan speed $c$ scales as $(q_{sat}-q_0)/H$ provided that the flux
be saturated at the crest.  This $1/H$ scaling of the speed was
initially proposed by Bagnolds \cite{B41}.  For typical values of the
shear velocity (table~\ref{NumericalValues}), $q_{sat}$ is of the
order of a few hundred $m^2/year$ which is the order of magnitude
found for $Q$ on the field (see section \ref{velocc} and in particular
equation~(\ref{c=Q/H})).  Regarding the sand supply dependence, there
has been no estimate of $q_0$ so far.  Field measurements show that
Bagnolds scaling for the speed is imperfectly verified, a better
approximation being $c \simeq Q/(H_0+H)$.  The existence of the
cut-off scale $H_0$ could come from the fact that the flux at the
crest is not totally saturated but depends on the ratio of the dune
length $L$ to the saturation length $l_{sat}$.
\bfig[t!]  \bc \epsfxsize=\linewidth \epsfbox{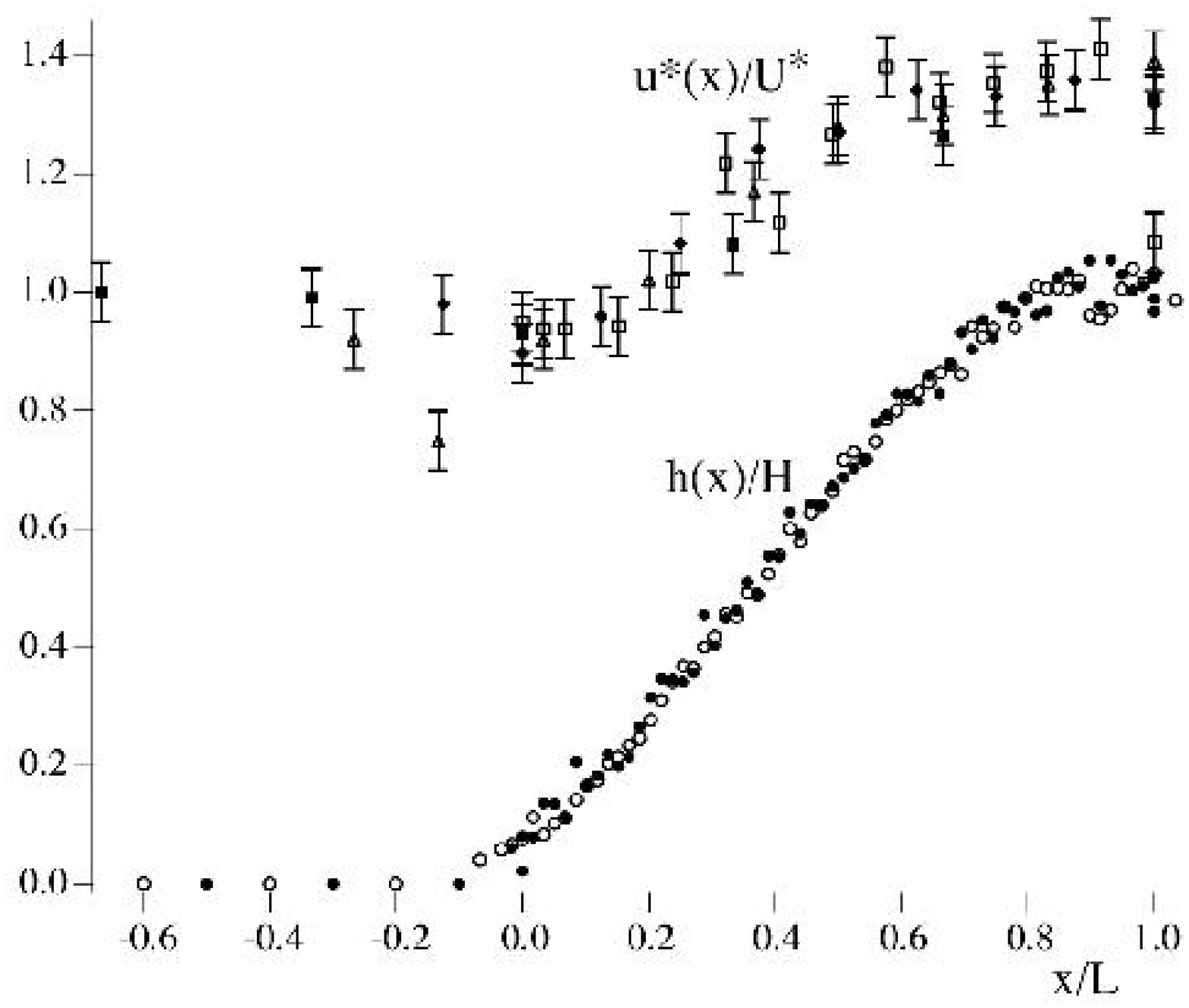}
\caption{The white and black circles are measurements of the central
profile of a barchan ($H=2.5~m$ and $L=36~m$) in the Negrita beach
(south Morocco).  The position $x$ is rescaled by the dune length $L$
and the profile $h(x)$ by the dune height $H$.  The shear velocity
$u_*$ on this dune is shown by white squares, rescaled by the shear
velocity $U_*$ far from the dune.  Three others shear velocity
profiles are shown, corresponding to barchans of various heights: the
black squares ($H=34~m$ and $L=200~m$) have been measured by Sauermann
{\it et al.} \cite{S01} in Jericoacoara (Brazil), the white triangles
($H=10~m$ and $L=100~m$) by Wiggs \cite{CWG93} in the Sultanate of
Oman, and the black diamonds ($H=5.5~m$ and $L=60~m$) by Howard {\it
et al.} \cite{HMGP78}.}
\label{SpeedUp}
\ec
\efig

\subsubsection{The wind shape relationship}
By the erosion/accretion process, the wind modifies the topography.
But in turn the topography modifies the wind.  The wind velocity has
been measured on the field around barchans of heights ranging from
$2.5~m$ to $34~m$ (figure~\ref{windvelocity}).  It turns out that the
it is almost independent of this height.  Even if the vertical
variation of the velocity is not perfectly logarithmic \cite{M88}, the
shear velocity can be computed at different place on the dune.  At the
edge of the dune, $u_*$ is observed to be slightly smaller than its
value $U_*$ far from the dune ($u_*/U_* \simeq 0.9$).  It then
increases along the dune back: at the inflexion point, $u_*/U_*$ is
around $1.2$ and it reaches $1.3$ to $1.4$ at the top of the dune.
\bfig[t!]  \bc \epsfxsize=\linewidth \epsfbox{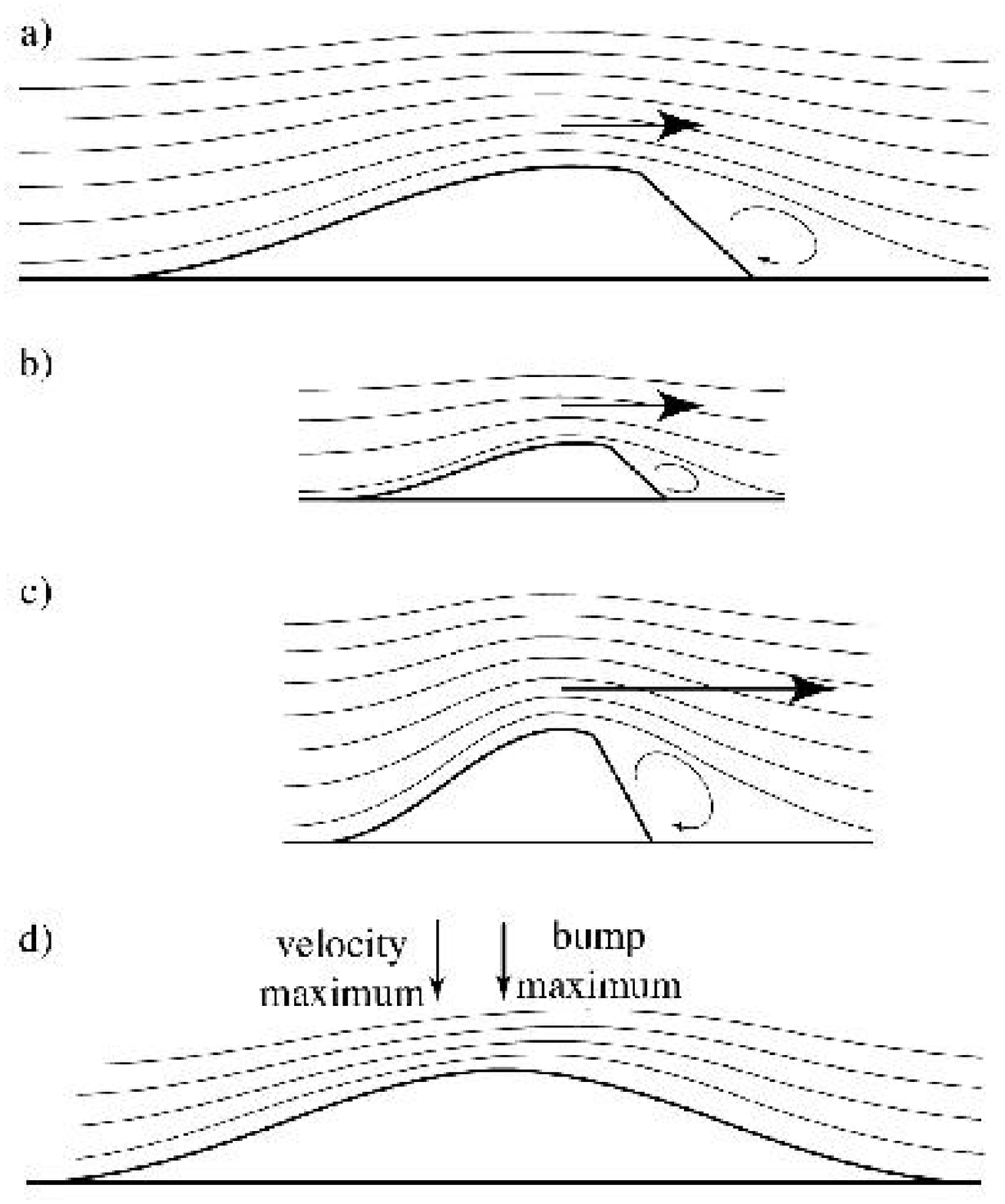}
\caption{A uniform ideal turbulent flow does not have any lengthscale,
meaning that the velocity at the top of a bump (a) is the same if the
bump is twice smaller (b).  (c) Due to a pressure effect, the velocity
at the top is larger if the bump length is divided by two but not the
height.  Even if the bump presents a symmetry downwind/upwind (d), the
velocity field does not present the same symmetry: the velocity is
larger before the crest than after so that the velocity maximum is
displaced upwind with respect to the bump maximum.}
\label{windvelocity}
\ec
\efig

The main theoretical problem is thus to determine the flow around a
given sand bump and in particular the lee recirculations.  The scaling
laws of the wind velocity field are somehow simpler to determine than
its spatial variations.  For a fully turbulent flow, the Reynolds
number $u_* H/\nu$ based on the dune size, is typically of the order
of $10^6$, and therefore viscous effects are negligible in front of
inertial effects.  This means that the whole velocity field is
proportional to the shear velocity far from the dune, noted $U_*$.

Let us first consider the -- theoretical -- case of a vertically
uniform turbulent flow.  An homogeneous turbulent flow does not have
any proper lengthscale.  The whole flow is thus scale invariant: two
dunes of the same shape but of different sizes should be surrounded by
the same velocity field (in rescaled coordinates).  To say it in crude
way, if the dune size is multiplied by two, the velocity at the top
remains the same (figure~\ref{windvelocity}).

The main effect is related to changes of pressure \cite{ZJ87,BAD99},
which is a non local function of the shape.  In other words, the
pressure and thus the velocity at some point of the dune is a function
of the whole dune shape.  Denoting by $h(x)$ the local height of a
two-dimensional dune (as that shown on figure~\ref{windvelocity}), we
can conclude from the previous considerations that the shear velocity
$u_*$ depends only on the slope $\ddx h$, which is the only
dimensionless field describing the dune shape.  But it is a non local
function of $\ddx h$.  Consider the dune shown in
figure~\ref{windvelocity} (a) and shrink it horizontally by a factor
of two (figure~\ref{windvelocity} c), the wind will be larger.  This
means that the velocity is sensitive to the aspect ratio of the bump,
to a kind curvature rescaled by the size of the dune.  This pressure
effect is a linear effect and exists even in the limit of small bumps.

If the direction of the wind is a symmetry axis for the dune, the wind
is symmetrical as well.  But if the dune admits a symmetry axis
perpendicular to the wind (figure~\ref{windvelocity} d), this symmetry
upstream/downstream is broken by the velocity field, as in the
classical case of the flow around a sphere.  This corresponds to an
irreversible feature of the flow: the streamlines are different when
the flow is reversed.  This asymmetry upwind/downwind is a non-linear
effect of the streamlines curvature on turbulence \cite{BAD99}.  It is
an inertial effect which, as previously, is scale invariant.

If the backward face is steep, the flow can no longer follow the form
of the dune.  The boundary layer separates from the bed and reattaches
down wind, enclosing a separation bubble.  Inside this bubble, the
wind flows back toward the obstacle so that the velocity profile is no
longer logarithmic.  Note that even the exact conditions under which
flow separates are not clear yet.  The best idea to model flow
separation has been proposed by Zeman and Jensen \cite{JZ85}.  They
suggested that the flow outside the recirculation bubble was precisely
that would have been observed if the streamlines separating the
recirculating flow from the turbulent boundary layer were solid.
Again, this idea of a dune envelope seams correct to the first order
but should be corrected by two effects.  First, it is somehow
arbitrary to prolong the dune by a surface of same roughness.  Second,
the separation surface is well defined on the average but fluctuates
in time while the actual sand bed does not.

To summarise, the velocity field is scale invariant in the ideal case
of a vertically uniform turbulent flow.  The velocity increases in
regions where the curvature is negative.  In the case of a symmetric
bump, the velocity field around this bump does not present the same
symmetry: the velocity is larger before the crest than after so that
the velocity maximum is displaced upwind with respect to the bump
maximum (figure~\ref{windvelocity} d).  This means that the velocity
is also sensitive to the local slope.

All these properties are present in the expression given by Jackson
and Hunt \cite{JH75,HLR88,WHCWWLC91} and simplified by Kroy {\it et
al.} \cite{KSH01} for the flow around a smooth flat hill.  In two
dimensions, it reads:
\be
\label{JacHunt}
u_*^2(x)=U_*^2\left(1 + A \! \int \! \frac{ds}{\pi s} \,\ddx h(x-s)
                          + B \ddx h(x) \right)
\ee
where $A$ and $B$ are almost constant (see below).  Jackson and Hunt
\cite{JH75}have predicted the values of $A$ and $B$ in the limit of a
vanishing aspect ratio $H/L$ -- at least smaller than $0.05$ but
equation~\ref{JacHunt} can also be used for dunes, introducing
effective coefficients $A$ and $B$.  The whole field $u_*(x)$ is
proportional to $U_*$ which is the velocity above a flat bed ($\ddx
h=0$).  It does not depend on the overall bump size since it is a --
non local -- function of the slope, only.  It is modulated by the
local slope (the $B$ term) and by a `dimensionless curvature' (the $A$
term) which takes the form of a convolution of the slope by the kernel
$1/x$.  Equation (\ref{JacHunt}), used on a dune prolonged by the
separation bubble modelled empirically, is the best known analytical
model of wind above a barchan.  It has for instance been used by Weng
{\it et al.} \cite{WHCWWLC91} who computed the erosion rate of a
barchan and by Kroy {\it et al.} \cite{KSH01} who integrated
numerically a complete model of dunes.

In the case of a turbulent boundary layer, there is actually a
characteristic lengthscale, the soil roughness $z_0$, which breaks the
scale invariance.  The previous effects remains valid but the fact
that the velocity profile depends logarithmically on height leads to
logarithmic corrections in the dune length to roughness ratio
\cite{JH75,ZJ87,HLR88,WHCWWLC91,BAD99}.  For instance, in the original
Jackson and Hunt model, the asymmetry upwind/downwind was directly
related to the soil roughness ($B\propto1/\ln(L/z_0)$).  We see on
this example that the effect of the roughness $z_0$ is not negligible:
$\ln(L/z_0)$ is of the order of $10$ for dunes.  But it also shows
that it is almost independent of the dune size: a factor $10$ on $L$
leads to a variation of $20\%$ of $\ln(L/z_0)$.
As a conclusion, the wind scale invariance is a very good
approximation, even with roughness effects.  This analysis is
confirmed by the wind measurements presented on figure~\ref{SpeedUp}.

\subsection{Avalanches}
\label{Avalanches}
\begin{figure*}
\epsfxsize=\linewidth
\epsfbox{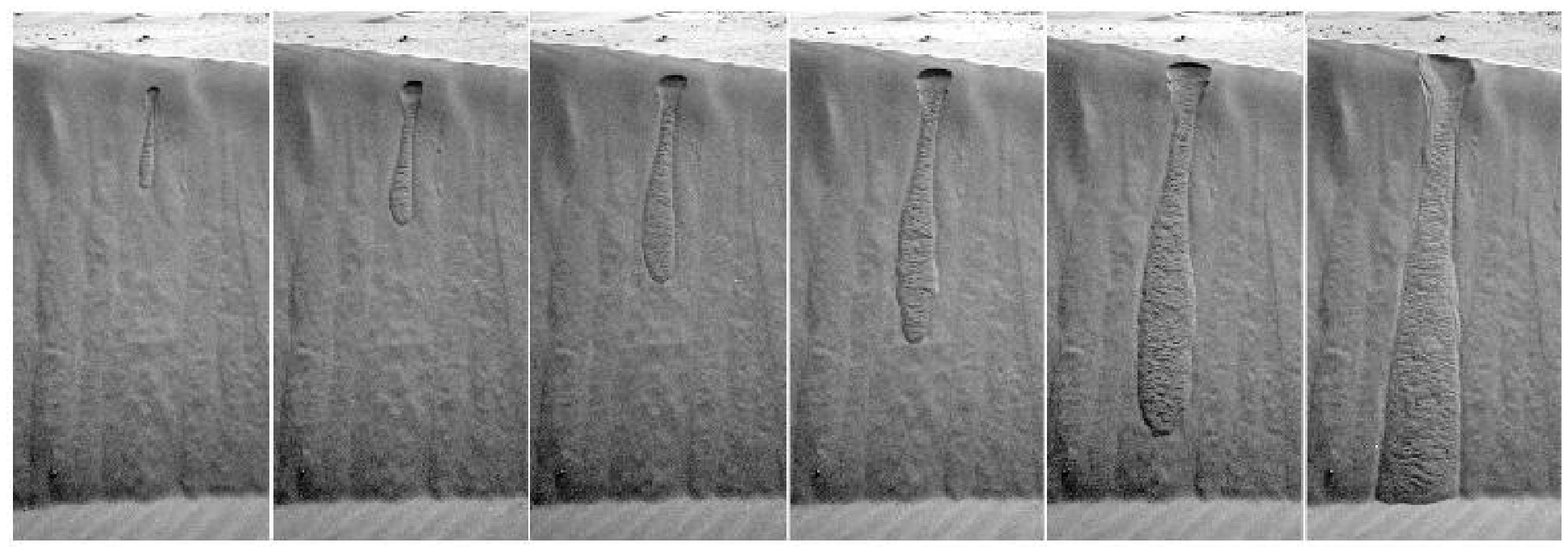}
\caption{When an avalanche spontaneously nucleates, usually in the
middle of the deposition zone, it both propagates downhill and uphill
and stops when the local slope has decreased below the dynamical
angle.
\label{Avalanche}}
\end{figure*}

The last important phenomenon takes place at the slip face.  If the
sand flux is not saturated upwind the dune, the back of the dune is
eroded and the sand flux increases.  This sand is deposed soon after
the brink, on the slip face, and forms a kind of snowdrift.  As shown
on figure~\ref{Avalanche} when the slope becomes locally larger than
the static friction coefficient $\mu_{s}$, an avalanche spontaneously
nucleates which propagates downward the slip face.  It is a dense flow
in which grains always remain in contact and which is limited to a
thin layer at the surface of the slip face.  As in solid friction
\cite{C1773}, this flow stops roughly when the slope as decreased
below the dynamical friction coefficient $\mu_{d}$.  The modelling of
avalanches have recently received pretty high attention from the
physicists \cite{BCRE94,BRDG98,P99b}.  In particular, we now have
accurate descriptions of granular surface flows by Saint-Venant
equations \cite{P99b,DAD99,AD00} governing the evolution of the free
surface, the flowing height and the mean velocity.

Fortunately, we do not need such refined models to understand
avalanches in the case of dunes.  Note first that the avalanche
duration (a few seconds to one minute) is always smaller than the time
separating avalanches (a few minutes to a few days) which is much
smaller than the turnover time.  Note also that the details of the
slip face do not react on the dune back since it is inside the
separation bubble.  Each individual avalanche propagates downward the
steepest slope and stops when the slope has decreased below $\mu_{d}$.
Thus, avalanches may be considered as an instantaneous slope
relaxation process which displaces sand along the steepest slope, when
the later is larger than $\mu_{d}$.

%______________________________________________________________________________
\section{Conclusion: open problems}
\label{problems}
\subsection{Field observations}
In the first part of this article, we have presented the field
observations about barchan dunes.  Let us give here a short summary.
A barchan is a crescentic dune propagating downwind on a firm soil.
When the direction of the wind is almost constant, these dunes can
maintain a nearly constant shape and size for very long times.  The
barchan mean velocity $c$ scales approximately with the inverse of its
height $H$: $c \simeq Q/(H_0+H)$.  It turns out that the flux $Q$
(typically $100~m^2/s$) and the height cut-off $H_0$ (of the order of
$1~m$) depend on time (probably through meteorological variations) and
on the dune field.  The barchan height $H$, its length $L$, its width
$W$ and the length of its horns $L_{wings}$ are, on the average,
related one to the others by linear relationships which depend on the
dune field but apparently not on the wind variations.  The typical
proportions of large dunes are $9$ for the ratio width to height, $6$
for the ratio length to height and $9$ for the ratio horn length to
height.  However, $H$, $L$ and $W$ are not proportional. It means that
barchans are not scale invariant.  The existence of a characteristic
size is confirmed by the fact that no barchan lower than $1~m$ are
observed.

The first thing to note is the strong dispersion of field
measurements: typically $50$ dunes have to be measured to establish
correctly one relation.  As a consequence, there is few reliable data
despite the numerous studies.  Moreover, the barchans morphology and
speed obviously depend on local parameters which have not been
identified so far: no systematic study has been made on the influence
of the wind speed, its fluctuations in direction, the nature of the
soil, the sand supply, the vicinity of other dunes, etc.  One of the
goal, in the future, will be the establishment of universal
relationships integrating the dependence on local parameters.  The
existence of a minimum size of dune also rises questions.  What
determines this size?  If a small conical sandpile quickly disappears
when eroded by the wind, how can barchan form?

These questions concern barchans in their individual behaviour.
Another class of problems concerns the global dynamics of dune fields
(ergs \cite{W72,W73}).  What determines the mean spacing between
dunes?  What selects the size of the dune?  What determines the sand
supply at the back of the dune and the leak by the horns?  Is sand
transport more efficient, on the average, in a barchan field when
compared to saltation over the desert floor?  What are the precise
conditions under which linear transverse dunes appear instead of a
barchan field?  These questions have received pretty low attention.
For instance there has been one measurement of crescent dune spacing
in Namibia by Lancaster \cite{L82} who have found a linear relation
between spacing and height.  But it was in a field in which the dunes
are very close one to the others.  Another example is provided by
Hastenrath \cite{H67,H87}, who have computed the histogram of dune
heights in the Pampa de La Joya (southern Peru).  In 1964 it exhibits
a sharp peak around $3.5~m$ and in 1983 around $1.5~m$.  These
measurements show that the dune fields are homogeneous in dune heights
but are not sufficient to determine the origin of the height
selection.  Concerning the average sand flux, the bulk transport i.e.
the transport of sand by the dunes have been measured by Lettau and
Lettau \cite{LL69} also in the Pampa de La Joya.  The total flux
between 1958 and 1964 was around $1~m^2/year$ which is a hundred times
smaller than the saturated flux on a flat soil: bulk transport is
apparently much less efficient than saltation over the flat soil.  If
the wind is able to transport much more sand in saltation, why does
the flux saturates in the places where barchans form?  How and where
can a dune field form?  Moreover they have shown that the mean flux
increases downwind meaning that the soil is eroded (at a rate $200~\mu
m/year$).  This article is the only tentative of description of sand
fluxes at the scale of the dune field.

As a conclusion, further studies will have to focus on the detailed
characterisation of barchan fields, aiming to get sufficient
statistics, as for the Pampa de La Joya.

\subsection{Dynamical mechanisms and dune modelling}
In the present state of the art, most of the dynamical mechanisms
important for barchans formation and propagation have been identified.
The explanation of dune propagation is simple: the back of the dune is
eroded by the wind and the sand transported in the air is deposed at
the brink and is redistributed on the slip-face by avalanches.  The
detailed description can be decomposed into two parts: the sand
transport over a sand bed for a given wind speed on the one hand and
the wind speed around a given bump on the other hand.

In this paper, we have proposed a coherent picture of sand transport.
If the wind strength is sufficiently large, grains can be directly
entrained by the wind.  They roll on the soil, take-off, fall back
again, rebound, are accelerated by the wind, fall back again and so
on.  Once the transport initiated, another mechanism of production
takes over the direct aerodynamic entrainment.  When the grains in
saltation collides the sand bed, they splash up a number of low energy
reptons which are accelerated by the wind and become saltons.  So, in
a first stage, the sand flux increases exponentially.  To accelerate
the grains, the wind has to give them some of its momentum.  Its
strength therefore decreases as the sand flux increases.  The
transport reaches equilibrium when the shear velocity has so decreased
that the number of reptons promoted to saltation just balance the
number of saltons remaining trapped after collisions with the sand
bed.  This saturation process takes time -- and space -- to establish.
The space lag between the edge of a sand sheet blown by the wind and
the point at which the flux saturates is given by the inertial length
$l_{drag}$ defined as the distance needed for a repton to become a
salton i.e.  to be accelerated to the wind velocity.  This saturation
length is the only relevant lengthscale of the problem and is directly
related to the minimum length of barchans.

There are still a few points concerning sand transport which need to
be clarified.  A first important aim is to reexamine experimentally
the difference between saltation and reptation, in particular the
controversial scaling of the saltation height and length, the
influence of a lateral or longitudinal slope, etc.  A second problem
is the anomalous scaling of the saturated flux, the apparent roughness
and probably other quantities, with the grain diameter.  This suggests
the existence of unknown mechanisms which would be worth studying.  A
third interesting study would be to reexamine experimentally the
saturation process, in particular the mechanisms and the scaling of
the saturation length.

The problem of the turbulent wind flow over a dune is different.  On
the one hand, there exist well known methods to solve this problem
numerically even though they take a lot of computation time.  On the
other hand, we know the basic principles: the velocity field is scale
invariant and is proportional to the wind velocity far from the
obstacle; it increases on a bump and is larger for an upwind than for
a downwind slope.  In middle, very few models have been proposed which
are both realistic and sufficiently simple to be understood.  The most
useful is certainly that of Jackson and Hunt \cite{JH75}.  The main
open problem is the description of the recirculation bubble which
requires either a full 3D simulation or very crude empirical
assumptions.

Theoretical and numerical studies of dunes are particularly
interesting and helpful to understand their dynamics.  The first aim
to compute numerically the rate at which sand is eroded or deposited
on barchan dunes goes back to Howard {\it et al.} \cite{HMGP78}.  To
do so, they used the topography of an actual barchan together with
laboratory measurements of the velocity field around a scale model of
this dune.  This work was completed by Weng {\it et al.}
\cite{WHCWWLC91} who also computed the erosion rate but this time,
using Jackson and Hunt approximation.  Wipperman and Gross \cite{WG86}
used a simpler model of flow over hills but computed the evolution in
time of a sandpile.  They were able to get a crescentic dune
propagating at a nearly constant velocity.  However, due to the
complexity of their model, this computation could only be performed on
one turnover time.  More recently, numerical models of the transient
of formation of two dimensional dunes were proposed by Stam \cite{S97}
and van Dijk {\it et al.} \cite{DAB99}.  Finally, Kroy {\it et al.}
\cite{KSH01} have proposed a complete model which takes into account
all the known mechanisms and which also leads to reasonable barchan
dunes.  In part 2, we will simplify this model and investigate
theoretically the shape and velocity of dunes.  Much simpler numerical
models such as cellular automatons \cite{NO93,W95,NY98} have been
proposed in the last decade which do not lead to realistic isolated
dunes but which are able to describe patterns of interacting dunes and
to test the influence of variable winds.  The next step is now to use
numerical simulations or laboratory experiments to supply for the lack
of control in field studies.  They will certainly help to find the
important parameters for the selection of dunes shape, size and
velocity but also to look at the interaction between dunes and even to
test the influence of complex wind regimes.

%______________________________________________________________________________
\begin{acknowledgement}
{\bf Acknowledgments\\} B. Andreotti and S. Douady would like to thank
F.~Naim, M.~Naim for their welcome and for the use of the Cemagref
wind tunnel.  The authors wish to thank P. Hersen and Y. Couder for
critical reading of the manuscript.  This work benefited from the
`Action Concert\'ee Incitative' of the french ministry of research.
\end{acknowledgement}

%______________________________________________________________________________

\end{document}